\documentclass{agujournal}
 
\usepackage{tikz}
\usepackage{dsfont}
\usepackage{bm}
\usepackage{amsmath}
\usepackage{amssymb}
\journalname{JGR-Earth Surface}

\begin{document}

\title{Unifying particle-based and continuum models of hillslope evolution with a probabilistic scaling technique}

%% ------------------------------------------------------------------------ %%
%
%  AUTHORS AND AFFILIATIONS
%
%% ------------------------------------------------------------------------ %%

% List authors by first name or initial followed by last name and
% separated by commas. Use \affil{} to number affiliations, and
% \thanks{} for author notes.  
% Additional author notes should be indicated with \thanks{} (for
% example, for current addresses). 

% Example: \authors{A. B. Author\affil{1}\thanks{Current address, Antartica}, B. C. Author\affil{2,3}, and D. E.
% Author\affil{3,4}\thanks{Also funded by Monsanto.}}

\authors{J. Calvert\affil{1}, M. Bal{\'a}zs\affil{2}, K. Michaelides\affil{3,4}}

\affiliation{1}{Department of Statistics, University of California, Berkeley, Evans Hall, Berkeley, CA, 94709, USA}
\affiliation{2}{School of Mathematics, University of Bristol, University Walk, Bristol, BS8 1TW, UK}
\affiliation{3}{School of Geographical Sciences, University of Bristol, University Road, Bristol, BS8 1SS, UK}
\affiliation{4}{Earth Research Institute, University of California, Santa Barbara, Santa Barbara, CA, 91306, USA}

%% \affiliation{4}{Fourth Affiliation}

%% Corresponding Author:
% Corresponding author mailing address and e-mail address:

% (include name and email addresses of the corresponding author.  More
% than one corresponding author is allowed in this LaTeX file and for
% publication; but only one corresponding author is allowed in our
% editorial system.)  

% Example: \correspondingauthor{First and Last Name}{email@address.edu}
\correspondingauthor{J. Calvert}{jacob\_calvert@berkeley.edu}

%% Keypoints, final entry on title page.

% Example: 
% \begin{keypoints}
% \item	List up to three key points (at least one is required)
% \item	Key Points summarize the main points and conclusions of the article
% \item	Each must be 100 characters or less with no special characters or punctuation 
% \end{keypoints}

%  List up to three key points (at least one is required)
%  Key Points summarize the main points and conclusions of the article
%  Each must be 100 characters or less with no special characters or punctuation 

\begin{keypoints}
\item Particle-based model of hillslope evolution
\item Probabilistic scaling of particle model gives continuum advection-diffusion equation
\item Bridges microscopic and macroscopic descriptions of hillslope evolution
\end{keypoints}

%% ------------------------------------------------------------------------ %%
%
%  ABSTRACT
%
%% ------------------------------------------------------------------------ %%

%% \begin{abstract} starts the second page 

\begin{abstract}
Relationships between sediment flux and geomorphic processes are combined with statements of mass conservation, in order to create continuum models of hillslope evolution. These models have parameters which can be calibrated using available topographical data. This contrasts the use of particle-based models, which may be more difficult to calibrate, but are simpler, easier to implement, and have the potential to provide insight into the statistics of grain motion. The realms of individual particles and the continuum, while disparate in geomorphological modeling, can be connected using scaling techniques commonly employed in probability theory. Here, we motivate the choice of a particle-based model of hillslope evolution, whose stationary distributions we characterize. We then provide a heuristic scaling argument, which identifies a candidate for their continuum limit. By simulating instances of the particle model, we obtain equilibrium hillslope profiles and probe their response to perturbations. These results provide a proof-of-concept in the unification of microscopic and macroscopic descriptions of hillslope evolution through probabilistic techniques, and simplify the study of hillslope response to external influences.
\end{abstract}

%% ------------------------------------------------------------------------ %%
%
%  TEXT
%
%% ------------------------------------------------------------------------ %%

\section{Introduction}

Hillslopes evolve topographically through a variety of erosional mechanisms ranging from slow diffusive processes (e.g. soil creep), to fast, localized processes (e.g. landslides). Over short timescales ($10^0$ - $10^1$ yr), hillslope sediment transport determines the redistribution of sediment and its delivery to the slope base. Over long timescales ($10^2$ - $10^5$ yr) the balance between, and integral of, individual erosional events determines the topographic form of hillslopes. Where advective processes dominate, hillslopes tend to be concave up, and where diffusive processes are more pronounced hillslopes become convex (e.g. \citep{carson1972hillslope,kirkby1971}). It is well-acknowledged that the processes shaping landscapes are inherently dynamic and stochastic \citep{dietrich2003geomorphic,roering2004soil,tucker2010modelling}, yet landscape evolution model (LEM) characterization of hillslope processes is often based on geomorphic transport laws (GTLs), mathematical formulations expressing erosion as an averaged process operating over long timescales \citep{dietrich2003geomorphic}. This discrepancy gives rise to a mathematical disconnect between the stochastic processes operating at the grain scale over the short term, and the evolution of hillslope topography over the long term.

In this paper we demonstrate a principled probabilistic scaling argument by which a particle-based description of hillslope sediment transport can be scaled to a continuum one representing long-term hillslope evolution. In other words, we present a mathematical argument for deriving a continuum description of hillslope erosion while remaining faithful to the particle-scale dynamics that operate over short time and space scales. 

GTLs are a compromise between a comprehensive physics-based description, which may be too complex to be parametrized through field observation, and rules-based modeling, which may lack a testable mechanistic footing \citep{dietrich2003geomorphic}. LEMs typically consist of a statement of mass conservation, GTLs for describing sediment transport in the form of differential equations, and numerical methods to approximate solutions to the GTLs \citep{tucker2010modelling}. Despite inherent simplifying assumptions associated with this approach, GTLs have been successful in simulating landform development in some environments, particularly associated with diffusive processes like creep and bioturbation (e.g. \citep{roering1999evidence}).

Particle-based models, which display a rich range of behavior despite their simplicity and ease of implementation, are an important alternative to this prescription of landscape evolution modeling \citep{tucker2010trouble,kessler2003self,davies2011discrete}. Traditionally, particle models have been criticized for using ``ad-hoc'' evolution rules and experimentally inaccessible parameters, and for neglecting the underlying transport physics \citep{dietrich2003geomorphic}. Accordingly, as continuum models have long been numerically implementable and experimentally verifiable, the use of GTLs has dominated studies of landscape evolution. However, the experimental validation of particle-based models is now possible using techniques for tracking grain motion \citep{mcnamara2004observations,habersack2001radio,roering2004soil,fathel2015experimental,roseberry2012probabilistic}. This, combined with their ability to incorporate particle mechanics and motion statistics, leads \citet{tucker2010trouble} to argue that particle-based models are no less fundamental than GTLs and should be used to complement continuum models.

While the case against the use of particle-based models has been undermined by experimental innovation, it is the theoretical development of nonlocal transport on hillslopes which best underscores the case for their use. Continuum models like those of \citet{culling1963soil} and \citet{andrews1987fitting} rely on locality assumptions, the assumption that sediment transport at position $x$ on a slope is a function of the hillslope conditions at $x$ (i.e. local land-surface slope) \citep{furbish2013sediment}. Locality assumptions are only valid when hillslope particles move short distances relative to the hillslope length \citep{tucker2010trouble}. Examples of local transport mechanisms are soil creep \citep{furbish2009statistical}, rainsplash \citep{dunne2010rain,furbish2009rain}, bioturbation and tree throw \citep{gabet2000gopher,gabet2003effects}. Nonlocal transport occurs when sediment transport at position $x$ depends on the hillslope characteristics a significant distance upslope or downslope of position $x$ \citep{furbish2013sediment} such that occurs in sheetwash sediment transport \citep{michaelides2012sediment,michaelides2014impact} and dry ravel \citep{gabet2012particle} on steep slopes. Accordingly, formulations of nonlocal transport must specify the relationship between flux and relative upslope or downslope, ultimately leading to assumptions on the distribution of particle travel distances \citep{furbish2010divots,furbish2013sediment} or the fitting of a fractional derivative operator \citep{foufoula2010nonlocal}. However, such relationships change as hillslopes evolve, and so particle-based approaches may be more appropriate \citep{gabet2012particle,dibiase2017slope}.

\begin{figure}[htbp]
\centering
\begin{tikzpicture}
    \node[anchor=south west,inner sep=0] (image) at (0,0) {\includegraphics[width=0.9\textwidth]{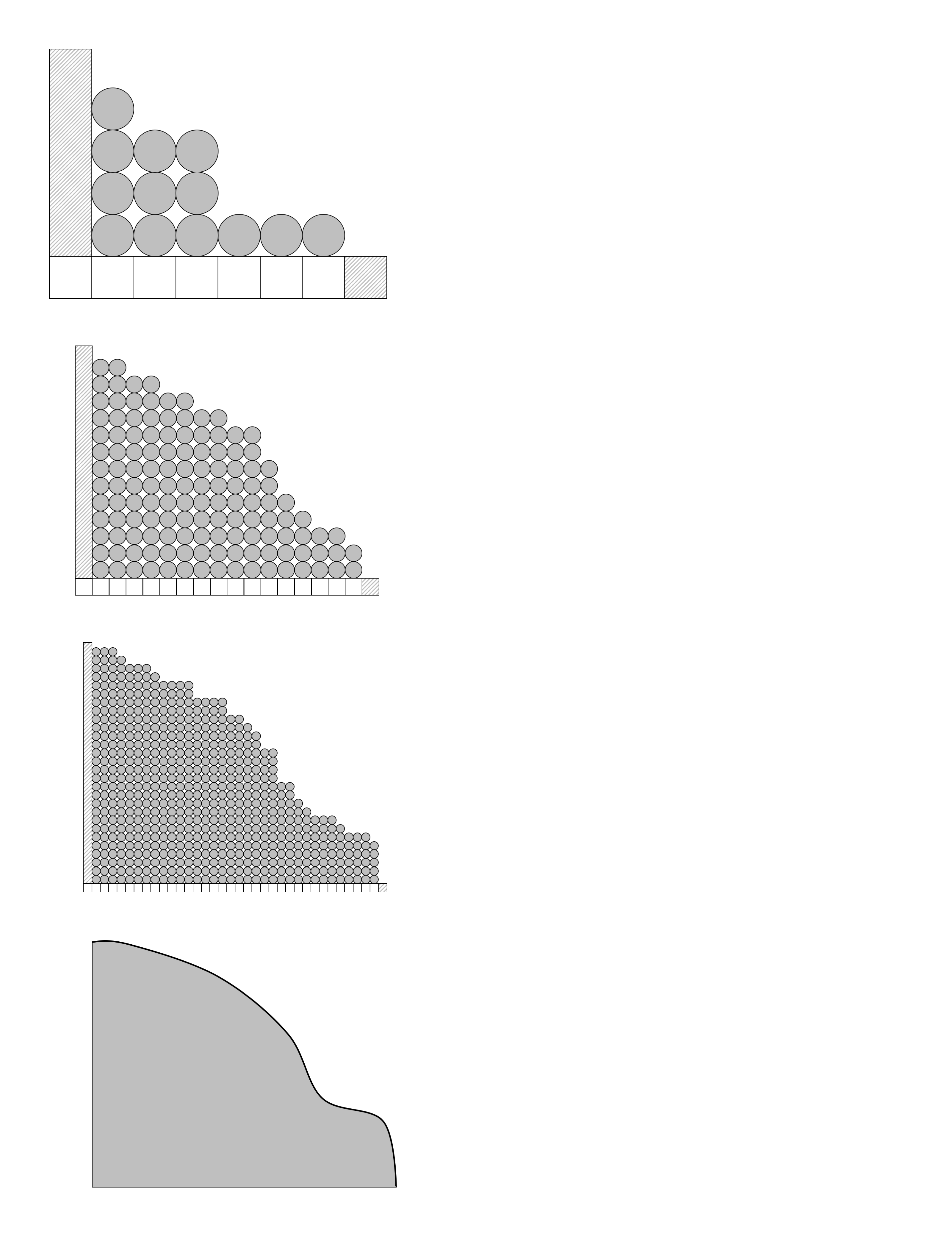}};
    \begin{scope}[x={(image.south east)},y={(image.north west)}]
        \node at (0.21,0.78) {\LARGE{\textbf{$i$}}};
        \node at (0.25,0.02) {\LARGE{\textbf{$x$}}};
   
       \node at (0.02,0.87) {\huge{\textbf{$h$}}};
        %\node at (0.02,0.94) {\huge{\textbf{$h$}}};
        \node at (0.06,0.15) {\huge{\textbf{$h$}}};
        %\node at (0.06,0.23) {\huge{\textbf{$h$}}};
        
        \node at (1.01,0.76) {\huge{\textbf{$\tau$}}};
        %\node at (0.465,0.94) {\huge{\textbf{$h$}}\LARGE{\textbf{$(i)$}}};
        %\node at (0.465,0.87) {\huge{\textbf{$h$}}\LARGE{\textbf{$(i)$}}};
        \node at (0.537,0.87) {\huge{\textbf{$h$}}\LARGE{$(i)$}};
        %\draw[->] (0.53,0.96) -- (0.255,0.83);
        
        \node at (1.01,0.05) {\huge{\textbf{$t$}}};
       % \node at (0.46,0.23) {\huge{\textbf{$h$}}\LARGE{\textbf{$(x)$}}};
        %\node at (0.46,0.15) {\huge{\textbf{$h$}}\LARGE{\textbf{$(x)$}}};
        \node at (0.537,0.15) {\huge{\textbf{$h$}}\LARGE{$(x)$}};
        %\draw[->] (0.53,0.25) -- (0.25,0.22);
        
        \draw[thick,->] (0.59,0.76) -- (0.98,0.76);
	\draw[thick,->] (0.59,0.76) -- (0.59,.96);
	\draw[thick,->] (0.59,0.05) -- (0.98,0.05);
	\draw[thick,->] (0.59,0.05) -- (0.59,0.25);
	
	% For the hillslope curve.
	\draw[thick] (0.095, 0.05) -- (0.095,0.25);
	\draw[thick] (0.095,0.05) -- (0.42,0.05);
	\draw[thick] (0.25,0.05) -- (0.25,0.04);
	\draw[dashed] (0.25,0.05) -- (0.25,0.21);
	
	% Label nodes at the top.
	\node[draw, align=center, text width=4cm] (A) at (0.205,1.02) {\large \textbf{Rescaling space ($i \mapsto x = \varepsilon i$)}};
	\node[draw, align=center, text width=4cm] (B) at (0.745,1.02) {\large \textbf{Rescaling~time ($\tau \mapsto t = \varepsilon^2 \tau$)}};
	\draw[thick] (A) -- (B);
	
	% Draw the limit arrow.
	%\node[rotate=270] at (0.05,0.50) {\Large{\textbf{$\xrightarrow{\makebox[3cm]{0 \rightarrow 0}}$}}};
	%\draw[thick,->] (0,0.75) -- (0,0.3) node[rotate=90,draw,midway, fill=white,text/rotate=90] {\Large{\textbf{$\varepsilon \rightarrow 0$}}};
%	\draw[thick,->] (0.48,1.02) -- (0.48,0.25) node[draw,midway, fill=white] {\Large{\textbf{$\varepsilon \rightarrow 0$}}};
	\draw[thick,->] (0.47,1.02) -- (0.47,0.25) node[draw,midway, fill=white] {\Large{\textbf{$\varepsilon \rightarrow 0$}}};
	
	% Put the plots here.
	 \draw[thick] (0.6,0.9) -- (0.65,0.9) -- (0.65,0.85) -- (0.7,0.85) -- (0.7,0.87) -- (0.75,0.87) -- (0.75,0.82) -- (0.8,0.82) -- (0.8,0.78) -- (0.85,0.78) -- (0.85,0.85) -- (0.9,0.85) -- (0.9,0.92) -- (0.95,0.92);
	 
	 \draw [thick] plot [smooth] coordinates { (0.6,0.17) (0.65,0.145) (0.7,0.135) (0.75,0.139) (0.775,0.12) (0.835,0.145)(0.885,0.18) (0.95,0.15) };
	
	% Breakout lines
	\draw[dashed] (0.6,0.9) -- (0.6,0.76);
	\draw[dashed] (0.95,0.92) -- (0.95,0.76);
	
	\draw[dashed] (0.6,0.745) -- (0.6,0.65);
	\draw[dashed] (0.95,0.745) -- (0.95,0.65);
	
	% Diagonals
	\draw[dashed] (0.95,0.65) -- (0.780,0.25);
	\draw[dashed] (0.6,0.65) -- (0.77,0.25);
	
	\draw[dashed] (0.78,0.25) -- (0.78,0.05);
	\draw[dashed] (0.77,0.25) -- (0.77,0.05);
	
	\draw[thick] (0.6,0.76) -- (0.6,0.75);
	\draw[thick] (0.95,0.76) -- (0.95,0.75);
	
	\draw[thick] (0.78,0.05) -- (0.78,0.04);
	\draw[thick] (0.77,0.05) -- (0.77,0.04);
	
	%\node at (0,0.945) {\huge{\textbf{A}}};
        %\node at (0.03,0.235) {\huge{\textbf{B}}};
	\node at (0,0.96) {\huge{\textbf{A}}};
        \node at (0.045,0.25) {\huge{\textbf{B}}};
        %\node at (1.01,0.96) {\huge{\textbf{C}}};
        %\node at (1.01,0.25) {\huge{\textbf{D}}};
        %\node at (0.99,0.96) {\huge{\textbf{C}}};
        %\node at (0.99,0.25) {\huge{\textbf{D}}};
        \node at (0.54,0.96) {\huge{\textbf{C}}};
        \node at (0.54,0.25) {\huge{\textbf{D}}};
        
        %\draw[help lines,xstep=.05,ystep=.05] (0,0) grid (1,1);
        %\foreach \x in {0,1,...,9} { \node [anchor=north] at (\x/10,0) {0.\x}; }
        %\foreach \y in {0,1,...,9} { \node [anchor=east] at (0,\y/10) {0.\y}; }
    \end{scope}
\end{tikzpicture}
\caption[Schematic of space and time rescaling.]{Schematic of space and time rescaling. Discrete space in a particle model of a hillslope, indexed by $i$ (\textbf{A}), is rescaled by a small parameter $\varepsilon$. In the limit as $\varepsilon$ approaches $0$, discrete space becomes continuous; accordingly, we replace $i$ with a continuous quantity $x$ = $\varepsilon i$ (\textbf{B}). After the rescaling, particles originally spaced by unit distance are spaced by $\varepsilon$. Consider instead the hillslope height at a particular site $i$, which changes in response to particle movements occurring on a timescale $\tau$ (\textbf{C}). After the rescaling of space, changes in hillslope height on timescale $\tau$ are too small to be observed, so the dynamics must be quickened by rescaling $\tau$ to $t$ with $\varepsilon^2$. Rescaling both space and time results in a macroscopic height $h (x)$ evolving on timescale $t$ (\textbf{D}).}
\label{fig:scaling_schematic}
\end{figure}
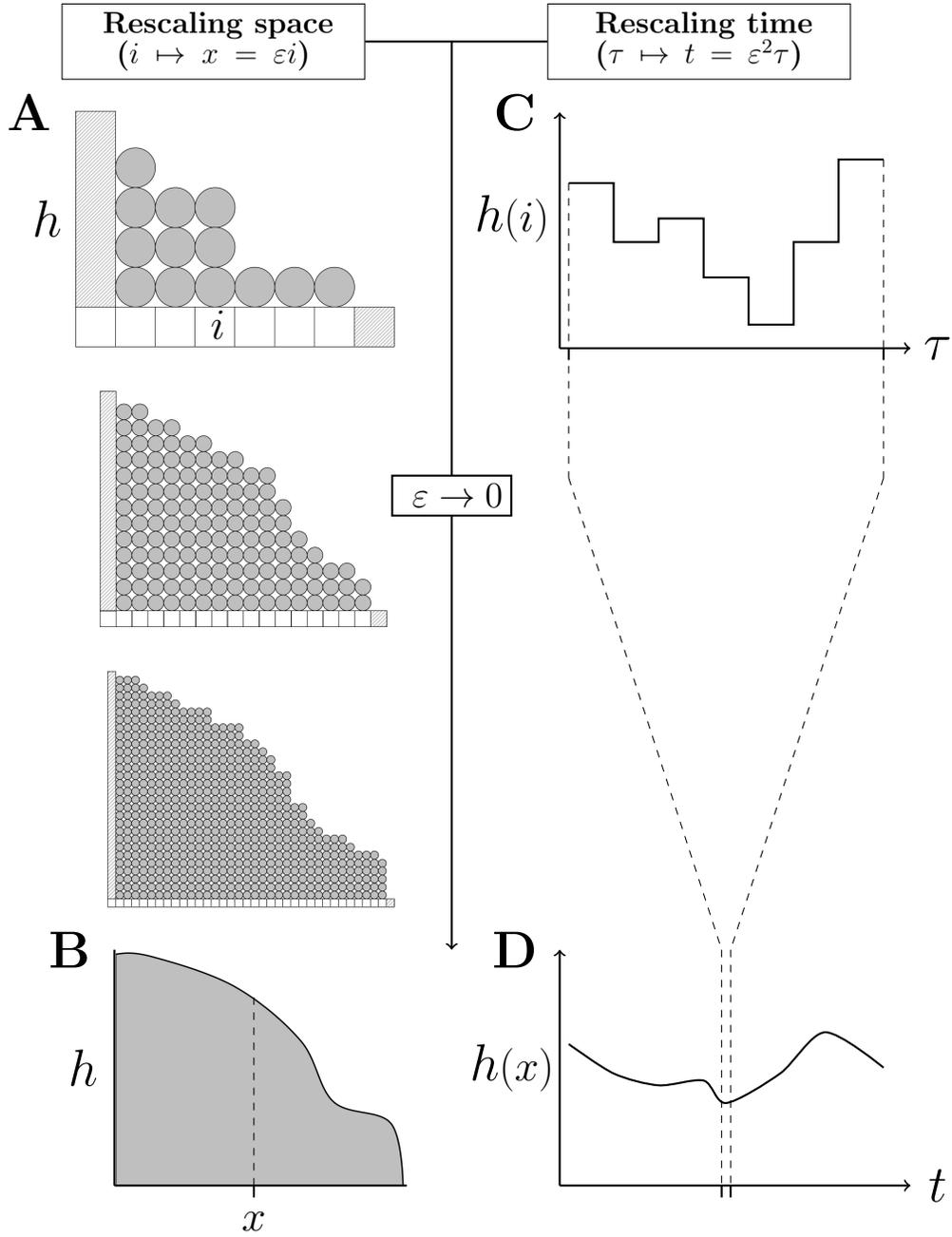

In order to effectively combine their strengths, the particle model must correspond, in some sense, to the continuum description. However, as \citet{tucker2010trouble} indicate, it is not clear how to identify such a pair. %, and previous attempts have been limited to qualitative comparisons between the particle and continuum descriptions \citep{furbish2010divots}. 
Indeed, referring to the particle-based models of \citet{tucker2010trouble} and \citet{gabet2012particle}, \citet{ancey2015stochastic} observe, ``there is no technique for deriving continuum equations from the rules used to describe particle behavior in this environment.'' Here, we demonstrate a principled \textit{probabilistic scaling} argument by which a particle-based description can be scaled to a continuum one with the two descriptions corresponding to one another. The probabilistic scaling procedure, illustrated in Figure~\ref{fig:scaling_schematic}, consists of scaling space and time by a small parameter, ultimately converting the microscopic evolution rules into a partial differential equation governing the macroscopic observables \citep{kipnis1999scaling,olla1993hydrodynamical,bahadoran2010strong}.

In Sections~\ref{sec:model} and \ref{sec:hydro}, we introduce a simple particle-based model of hillslope evolution and provide a heuristic scaling argument, which identifies a corresponding continuum description in the form of an advection-diffusion equation. Critically, the particles of our model correspond to units of hillslope \textit{gradient}, not hillslope height. This element of indirection softens the distinction between local and nonlocal transport and, for this reason, our model can represent diverse geomorphic processes and the scaling argument applies uniformly across various transport regimes.

Finessing nonlocal transport through indirection comes at the expense of immediate access to information about particle hopping distances and fluxes. This contrasts the convolutional approaches of \citet{foufoula2010nonlocal} and \citet{furbish2010divots}, which express sediment flux arising from nonlocal transport as an integral over relative upslope. While such methods enable detailed calculation of fluxes, they require as input assumptions about the distribution of particle hopping distances and hillslope topography \citep{gabet2012particle,furbish2010divots,furbish2013sediment}. When these detailed outputs are unnecessary, the requisite inputs are unavailable, or corresponding simulations are computationally expensive, a particle-based approach may be preferable.

Section~\ref{sec:sim} describes simulations of the particle system for various choices of microscopic parameters, including both linear and nonlinear slope dependence, to exhibit the types of hillslope profiles which form and how fluxes arise in response to hillslope perturbations. Additionally, to translate simulation results into empirically testable predictions, we suggest a principled way of fitting model parameters from data and assigning dimensions to model outputs. Finally, in Section~\ref{sec:disc}, we discuss the relation of this paper to the hillslope evolution and nonlocal transport literature and suggest future work, which takes advantage of a dual, particle-based and continuum approach.

%%%%%%%%%%%%%%%%%
%% GARGANTUAN FIGURE %%
%%%%%%%%%%%%%%%%%

\section{A particle-based model of hillslope evolution}\label{sec:model}

\subsection{Specifying state space and dynamics}\label{sub:dynamics}

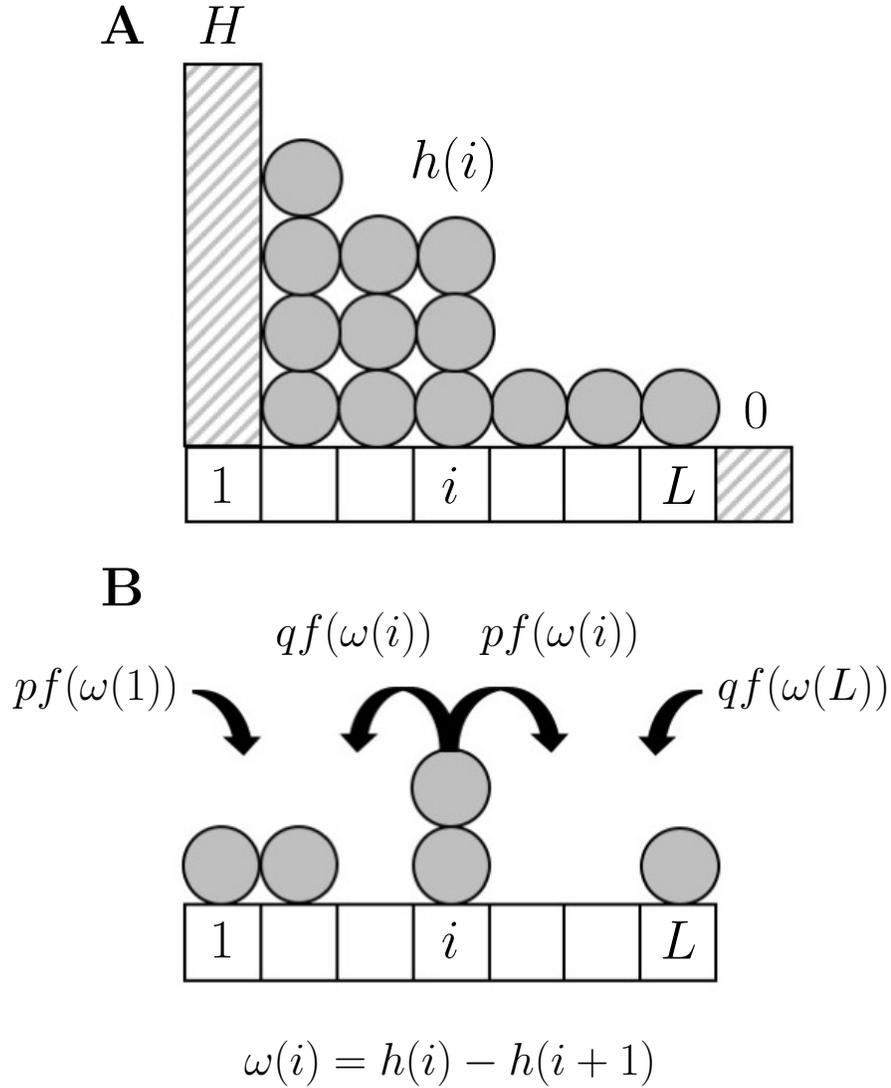
\begin{figure}[htbp]
\centering
\begin{tikzpicture}
    \node[anchor=south west,inner sep=0] (image) at (0,0) {\includegraphics[width=0.7\textwidth]{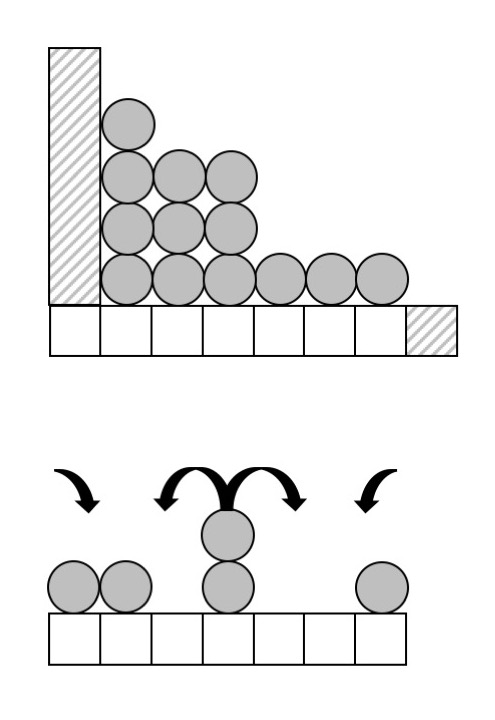}};
    \begin{scope}[x={(image.south east)},y={(image.north west)}]
        \node at (0.15,0.97) {\huge{\textbf{$H$}}};
        \node at (0.15,0.54) {\huge{\textbf{$1$}}};
        \node at (0.15,0.115) {\huge{\textbf{$1$}}};
        \node at (-0.02,0.35) {\LARGE{\textbf{$p f (\omega (1))$}}};
        \node at (0.94,0.35) {\LARGE{\textbf{$q f (\omega (L))$}}};
        %\node at (-0.01,0.35) {\LARGE{\textbf{$\alpha (\omega (1))$}}};
        %\node at (0.925,0.35) {\LARGE{\textbf{$\beta (\omega (L))$}}};
        \node at (0.33,0.40) {\LARGE{\textbf{$q f (\omega (i))$}}};
        \node at (0.61,0.40) {\LARGE{\textbf{$p f (\omega (i))$}}};
        \node at (0.46,0.54) {\huge{\textbf{$i$}}};
        \node at (0.46,0.115) {\huge{\textbf{$i$}}};
        \node at (0.77,0.115) {\huge{\textbf{$L$}}};
        \node at (0.77,0.54) {\huge{\textbf{$L$}}};
        \node at (0.875,0.61) {\huge{\textbf{$0$}}};
        \node at (0.465,0.84) {\huge{\textbf{$h (i)$}}};
        \node at (0.46,0.0) {\LARGE{\textbf{$\omega (i) = h (i) - h (i+1)$}}};
        
        \node at (0.015,0.97) {\huge{\textbf{A}}};
        \node at (0.015,0.445) {\huge{\textbf{B}}};
        
        %\draw[help lines,xstep=.1,ystep=.1] (0,0) grid (1,1);
        %\foreach \x in {0,1,...,9} { \node [anchor=north] at (\x/10,0) {0.\x}; }
        %\foreach \y in {0,1,...,9} { \node [anchor=east] at (0,\y/10) {0.\y}; }
    \end{scope}
\end{tikzpicture}
\caption[Mapping changes in a height profile to a gradient ZRP.]{Schematic of the mapping between the hillslope height (\textbf{A}) and corresponding hillslope gradients (\textbf{B}) of the particle model. The height of the hillslope's leftmost site ($i$ = $1$) is fixed at a height of $H$ and the rightmost site ($i=L+1$) is fixed at $0$ (\textbf{A}). In the gradient process (\textbf{B}), particles in the bulk ($1< i < L$) hop to the left and right with rates $qf(\omega (i))$ and $pf(\omega (i))$, respectively; particles at the left boundary move right at rate $pf(\omega (1))$ and those at the right boundary move left at rate $qf(\omega (L))$.}
\label{fig:fixedheightzrp}
\end{figure}

As our goal is to model hillslope profiles, we begin by considering a 1D grid of $L+1$ labeled sites, which each contain some number of identical ``units'' of hillslope (Figure~\ref{fig:fixedheightzrp}A). We fix the number of units at site $1$ to be $H$, and the number at site $L+1$ to be $0$. The process of hillslope evolution could then occur via the rearrangement of the units across sites $2$ to $L$, according to some dynamics. However, our analysis becomes simpler if we instead consider a corresponding ``gradient'' particle system, where the particles represent differences  in the number of hillslope hunks between adjacent grid sites (Figure~\ref{fig:fixedheightzrp}B). That is, if there are $h_\tau (i)$ units at site $i$ and time $\tau$ and $h_\tau (i+1)$ at site $i+1$, we place $\omega_\tau (i) = h_\tau (i) - h_\tau (i+1)$ particles at site $i$ of the gradient system. Note that, because we fixed site $L+1$ to have $0$ units, $\omega_\tau (L) = h_\tau (L)$. Additionally, because we fixed site $1$ to have $H$ units, $\sum_{i=1}^L \omega_\tau (i) = H$; we have conservation of gradient particles. In order to complete the specification of the gradient process, we need to describe the ways in which particles are allowed to move. 

Figure~\ref{fig:fixedheightzrp} summarizes the rules governing the dynamics. Particles hop after exponentially-distributed waiting times, with rates given as follows. For sites $i \neq 1,\,L$, a particle will hop $i \rightarrow i+1$ with rate $pf(\omega_\tau (i))$ and $i \rightarrow i-1$ with rate $qf(\omega_\tau (i))$, where $p, q \in (0,1)$ and $p+q = 1$, and $f(\omega_\tau (i))$ is a nondecreasing function of $\omega_\tau (i)$ with $f(0) = 0$. The requirement that $f$ be nondecreasing in $\omega_\tau (i)$ formalizes the intuition that the dynamics on steep slopes happen at least as quickly as those on gradual slopes. At the left boundary $i=1$, a particle hops $1 \rightarrow 2$ with rate $pf(\omega_\tau (1))$ and, at the right boundary $i=L$, a particle hops $L \rightarrow L-1$ with rate $qf (\omega_\tau (L))$. As the number of gradient particles, $\omega_\tau (i)$, represents the steepness of the hillslope at site $i$, a gradient particle hopping to site $i$ corresponds to the hillslope becoming steeper at $i$. In terms of hillslope profile $h_\tau$, this could reflect deposition at site $i$ or removal at site $i+1$, both of which would cause the hillslope to become steeper at $i$.

%Like $f$, $\alpha$ and $\beta$ are nondecreasing functions of their arguments, with $\alpha (0) = \beta (0) = 0$. At the left boundary $i=1$, a particle hops $1 \rightarrow 2$ with rate $\alpha (\omega_\tau (1))$ and, at the right boundary $i=L$, a particle hops $L \rightarrow L-1$ with rate $\beta (\omega_\tau (L))$. Like $f$, $\alpha$ and $\beta$ are nondecreasing functions of their arguments, with $\alpha (0) = \beta (0) = 0$. 

Our model is a type of \textit{continuous-time Markov process}, known in the statistical physics and probability literature as a ``zero-range process'' because particles hop at rates which depend on the occupancy of their current site. In this sense, there is a zero-range interaction between particles occupying the same site. Note that particles in the gradient process only hop unit distances, unlike particles in the model of \citet{tucker2010trouble}. While gradient particles redistribute locally, the corresponding changes in the original height profile need not be. 

\subsection{Identifying the stationary distributions of the particle model}\label{sec:stationary}

The stationary distributions of the gradient process are those probability distributions over occupancies $\omega (i)$ which are unchanged by the dynamics specified in Section \ref{sub:dynamics}. To find the stationary distributions, it suffices to enforce a ``detailed balance'' condition between configurations 
\begin{linenomath*}\[\omega = \{ \omega (1), \, \dots, \, \omega (i),\, \omega (i+1), \, \dots, \, \omega (L) \} \quad \text{and}\]\end{linenomath*} \begin{linenomath*}\[\omega^{i\rightarrow i+1} = \{ \omega (1), \, \dots, \, \omega (i) - 1,\, \omega (i+1) + 1, \, \dots, \, \omega (L) \},\]\end{linenomath*} which reads
\begin{linenomath*}\begin{equation}\label{eq:detailedbalance}
\mathds{P} (\omega) \cdot pf(\omega (i)) = \mathds{P} (\omega^{i\rightarrow i+1}) \cdot qf(\omega (i+1)+1).
\end{equation}\end{linenomath*} That is, in equilibrium, the frequency of moving from one configuration to a second is exactly balanced by the frequency of the reverse process. 

Surprisingly, despite the many interactions between particles, the probability distribution $\mathds{P}(\omega)$ of observing the particle system in configuration $\omega$ in equilibrium can be expressed as a product of decoupled marginal distributions for each site $\mathds{P}(\omega) = \prod_{i=1}^L {\mathds{P}_i}^{\theta_i} (\omega (i))$. Informally, at equilibrium, the probability of seeing a certain number of gradient particles at a site is independent of all other sites. This property enables us to study the simpler marginal distributions $\mathds{P}_i^{\theta_i}$ instead of $\mathds{P}$, and would not be present if we had instead modeled the hillslope directly, with particles representing units of hillslope height. The marginal distributions have the form
\begin{linenomath*}\begin{equation}\label{eq:dist}
{\mathds{P}_i}^{\theta_i} (\omega (i)) = \frac{e^{\theta_i \omega (i)}}{f(\omega (i))! \, Z(\theta_i) } \quad \theta_i \in \mathds{R},
\end{equation}\end{linenomath*} with $f(z)! = \prod_{k=1}^z f(k)$ and $f(0)! = 1$. $Z(\theta_i) = \sum_{k=0}^{\infty} e^{\theta_i k} / f(k)!$ is a normalization constant, which is assumed to be finite. In Appendix \ref{app:detailedbalance}, we show that Equation \ref{eq:dist} indeed satisfies the detailed balance condition of Equation \ref{eq:detailedbalance}, so long as $\exp (\theta_{i+1} - \theta_i) = p/q$. %In Appendix \ref{app:detailedbalance}, we show that Equation \ref{eq:dist} indeed satisfies the detailed balance condition of Equation \ref{eq:detailedbalance}, so long as (i) $\exp (\theta_{i+1} - \theta_i) = p/q$, (ii) $\alpha(\omega (1)) = pf(\omega (1))$, and (iii) $\beta(\omega (L)) = qf(\omega (L))$. 

Using the stationary distributions ${\mathds{P}_i}^{\theta_i}$, we would like to calculate the stationary density, that is, the expected number of gradient particles occupying each site in equilibrium. Technically, this quantity depends on the choice of hillslope height $H$, and so we should calculate the \textit{conditional} expected number of gradient particles at each site. For an arbitrary choice of $f(\omega (i))$, parameter $\theta_i$, and fixed height $H$, the density at a site $i$ is
\begin{linenomath*}\begin{equation}\label{eq:occupancy}
{\rho}^{\theta_i | H} (i) = \mathds{E}^{\theta_i} \Bigg( \omega (i) \, \Bigg| \sum_{j=1}^L \omega (j) = H \Bigg) = \sum_{k=0}^H k \cdot \mathds{P}^{\theta_i} \Bigg( \omega (i) = k \, \Bigg| \sum_{j=1}^L \omega (j) = H \Bigg), %= \sum_{\wi{i}=0}^H \wi{i} \cdot \frac{\mui{i} (\wi{i})\cdot \mathds{P}^{\theta} (\sum_{j\neq i} \wi{j} = H-\wi{i})}{\mathds{P}^{\theta} (\sum_{j = 1}^L \wi{j} = H)}, 
\end{equation}\end{linenomath*} where $\mathds{E}^{\theta_i}$ is the expectation with respect to the distribution $\mathds{P}^{\theta_i}$ and the notation $\Big| \sum \omega = H$ indicates conditioning on the sum of gradient particles being $H$. The sum over $k$ in \eqref{eq:occupancy} is an average over the numbers of gradient particles which could be at site $i$, with a weighting based on the probability of observing $k$ particles at site $i$, subject to the configuration having a total of $H$ gradient particles.

Note that this density describes the average number of particles at each site in equilibrium for the gradient process, not the original hillslope profile. In order to obtain a typical hillslope profile, the density must be inverted using $\omega (i) = h (i) - h (i+1)$, which leads to \begin{linenomath*}\begin{equation}\label{eq:inversion} h (i) = \sum_{j=i}^L \omega (j).\end{equation}\end{linenomath*}

\subsection{Hillslope profiles for linear rate}\label{sec:example}

We can calculate Equation \ref{eq:occupancy} explicitly for the choice of linear rate, $f(\omega (i)) = \omega (i)$, corresponding to the gradient particles hopping with rate proportional to local gradient. For this choice of $f(\omega (i))$, the stationary distributions are Poisson
\begin{linenomath*}\begin{equation}\label{eq:poisson}
{\mathds{P}_i}^{\theta_i} (\omega (i)) = \frac{e^{\theta_i\,\omega (i)}\cdot e^{-e^{\theta_i}}}{\omega (i)!}.
\end{equation}\end{linenomath*} In Appendix \ref{app:density}, we show that using (\ref{eq:poisson}) with (\ref{eq:occupancy}) gives
\begin{linenomath*}\begin{equation}\label{eq:density}
{\rho}^{\theta_i | H} (i) = H\frac{e^{\theta_i}}{\sum_{j=1}^L e^{\theta_j}} = H\cdot \Bigg(\frac{p}{q}\Bigg)^{i-1}\frac{\Big(\frac{p}{q}\Big) -1}{\Big(\frac{p}{q}\Big)^L - 1}.
\end{equation}\end{linenomath*} where the second equality follows from condition (i) on the $\theta_j$. 

\begin{sidewaysfigure}%[htbp]
\centering
\begin{tikzpicture}
    \node[anchor=south west,inner sep=0] (image) at (0,0) {\includegraphics[width=\textwidth]{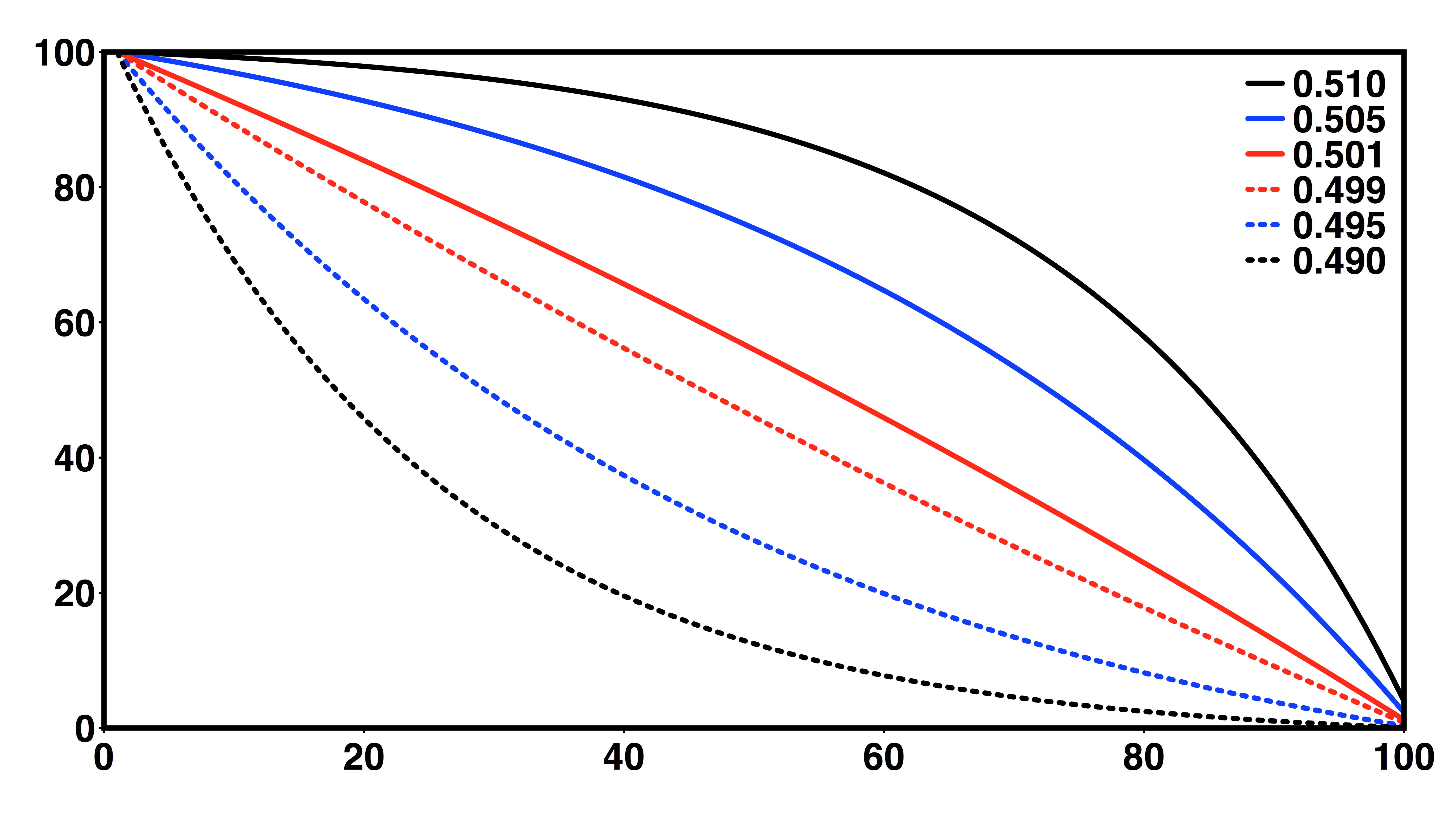}};
    \begin{scope}[x={(image.south east)},y={(image.north west)}]
        \node at (0.013,0.53) {\Huge{\textbf{$h (i)$}}};
        \node at (0.52,0.035) {\Huge{\textbf{$i$}}};
        \node[fill=white, opacity=0.35] at (0.615,0.72) {\Huge{\textbf{soil creep}}};
        \node[fill=white, opacity=0.35] at (0.51,0.55) {\Huge{\textbf{sheet wash}}};
         \node[fill=white, opacity=0.35] at (0.39,0.4) {\Huge{\textbf{sheet wash}}};
         \node[fill=white, opacity=0.35] at (0.39,0.32) {\Huge{\textbf{with rills/gullies}}};
        %\draw[thick,->] (0.7,1) -- (0.7,0.7) node[midway, fill=white, opacity=0.5] {\Large{\textbf{soil creep}}};
         %\draw[thick,->] (0.1,0.1) -- (0.35,0.35) node[] {\Large{\textbf{rivers}}};
        %\draw[help lines,xstep=.1,ystep=.1] (0,0) grid (1,1);
        %\foreach \x in {0,1,...,9} { \node [anchor=north] at (\x/10,0) {0.\x}; }
        %\foreach \y in {0,1,...,9} { \node [anchor=east] at (0,\y/10) {0.\y}; }
    \end{scope}
\end{tikzpicture}
\caption[Example hillslopes produced by the gradient ZRP.]{The hillslope profiles produced by the particle model for $f(\omega (i)) = \omega (i)$, fixed $H = L = 100$, and values of $p$ shown in the legend, where we fix $p+q = 1$. We indicate soil creep, sheet wash, and sheet wash with rills/gullies as geomorphic processes which could be modeled by these curves, in analogy with the characteristic-form profiles of \citet{kirkby1971}.}
\label{fig:exp_curves}
\end{sidewaysfigure}
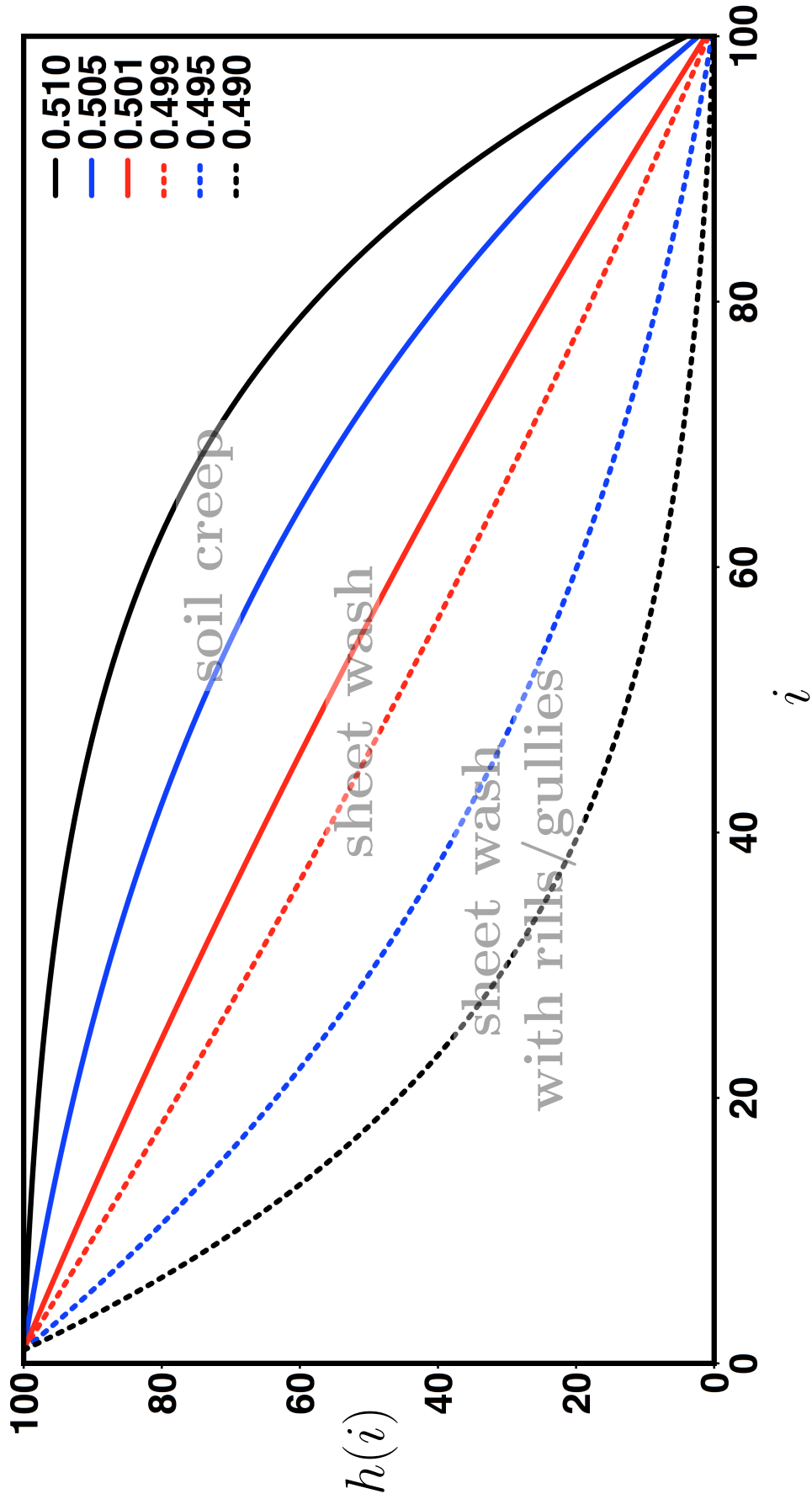

We can invert Equation~\ref{eq:density} with $h (i) = \sum_{j=i}^L \omega (j)$ to get the corresponding hillslope profile
\begin{linenomath*}\begin{equation}\label{eq:heights}
    h (i) = H\frac{\Big(\frac{p}{q}\Big)^i - \Big(\frac{p}{q}\Big)^{L+1}}{\Big(\frac{p}{q}\Big) - \Big(\frac{p}{q}\Big)^{L+1}} \quad \text{for} \quad 1 \leq i \leq L,
\end{equation}\end{linenomath*} which describes the expected hillslope profile in equilibrium. Examples of such profiles are provided for a range of $p/q$ values in Figure~\ref{fig:exp_curves}.

\subsection{Hillslope profiles for constant rate}\label{sec:stepexample}

We can also calculate $\rho^{\theta_i} (i) = \mathds{E}^{\theta_i} \omega (i)$, in absence of conditioning on $H$, for a choice of constant rate: $f(\omega (i)) = 1$ if $\omega (i) > 0 $ and $f(\omega (i)) = 0$ if $\omega (i) = 0 $. Whereas, in the case of linear rate, the dynamics depended on the local gradient, the constant rate case corresponds to a dynamics which evolves steep slopes at the same rate as gradual slopes. The fact that the conditioning matters little to the stationary hillslope profile follows from a large-deviations-type argument, which we omit here for brevity.

 The occupancies $\omega (i)$ are distributed as geometric random variables, that is \begin{linenomath*}\[
 \mathds{P_i}^\theta_i(\omega (i)) = \frac{e^{\theta_i \omega (i) }}{Z(\theta_i)f(\omega (i))!}=\frac{ e^{\theta_i \omega (i)}}{Z(\theta_i)} = e^{\theta_i \omega (i)} (1 - e^{\theta_i}),
\]\end{linenomath*}
thus the density of gradient particles has the following simple form
\begin{linenomath*}\begin{equation}\label{eq:steprho}
 \rho^{\theta_i} (i)=\frac{e^{\theta_i}}{1-e^{\theta_i}}, \end{equation}\end{linenomath*}
valid for $(\theta_i <0)$. The stationary distribution requires $e^{\theta_{i+1} - \theta_i} = p/q$, or
\begin{linenomath*}\[
 e^{\theta_i}=c\cdot\left(\frac {p}{q}\right)^i,\qquad c>0\,,\quad i<\frac{-\ln c}{\ln p-\ln q}
\]\end{linenomath*} which, combined with \eqref{eq:steprho}, gives the discrete gradient of the hillslope
\begin{linenomath*}\begin{equation}\label{eq:steprho2}
 \rho^{\theta_i} (i) = \frac{c\cdot\left(\frac {p}{q}\right)^i}{1-c\cdot\left(\frac pq\right)^i}.
\end{equation}\end{linenomath*} To obtain the expected hillslope profile corresponding to \eqref{eq:steprho2}, we apply $h(i) = \sum_{j=i}^L \omega (j)$ and substitute \eqref{eq:steprho2}, resulting in
\begin{linenomath*}\[
	h(i) = \sum_{j=i}^L \frac{c\cdot\left(\frac {p}{q}\right)^j}{1-c\cdot\left(\frac pq\right)^j}.
\]\end{linenomath*} We note that $c$ can be chosen to fit the left boundary condition for height $h(1) = H$.

\subsection{Particle model recap}\label{sec:particlemodelrecap}
We recall some key points from Section~\ref{sec:model} before proceeding to the scaling.\begin{enumerate}
\item The particles of the model represent units of slope, not units of height.
\item Particles move according to a rate function $f$ which is not necessarily linear.
\item To obtain a hillslope profile, the gradient particle profile must be summed according to \eqref{eq:inversion}.
\item Hillslope profiles can be calculated explicitly when $f$ is linear or constant; simulated otherwise.
\end{enumerate}

\section{Heuristic derivation of the continuum equation}\label{sec:hydro}

We now return to a general setting, where the form of $f(\omega_\tau (i))$ is unspecified, to identify the continuum equation corresponding to the particle-based model of Section~\ref{sec:stationary}. As in Section~\ref{sec:example}, the density $\rho_\tau (i) = \mathds{E}^{\theta_i} \omega_\tau (i)$ is the object of interest, the scaling of which wholly characterizes the gradient process in the limit of macroscopic time and space scales. We denote the particle model's time by $\tau$ and choose the scaling $t = \tau/dL^2$, $x = i/L$, with the interpretation that we ``zoom out'' by a factor of $L$ and speed up the process by a factor of $L^2$, in order to observe changes on the new spatial scale. This is the idea expressed in Figure~\ref{fig:scaling_schematic} with the small parameter $\varepsilon$ chosen in terms of the hillslope length as $\varepsilon = 1/L$, so $\varepsilon \rightarrow 0$ as $L \rightarrow \infty$. The time constant $d$ will become relevant in Section~\ref{sec:dimension}. We thus identify the rescaled density as \begin{linenomath*}\begin{equation*}\label{eq:rescaled_density} \varrho_t (x) := \mathds{E}^{\rho} \omega_{tdL^2} (xL), \end{equation*}\end{linenomath*} where the expectation with respect to $\rho$ is justified in Appendix~\ref{app:rho}.

We require that $p$ and $q$ become increasingly close in value when scaling $\rho_\tau (i)$. The intuition for this choice comes from the $f(\omega_\tau (i)) = \omega_\tau (i)$ curves of Figure~\ref{fig:exp_curves}, which indicate that increasing $p$ relative to $q$ results in a profile more closely resembling a step function. The scaling procedure will only serve to accentuate this resemblance and so, to avoid a degenerate rescaled density $\varrho_t (x)$, we choose the ``weakly asymmetric'' limit, where $p = \frac{1}{2} + \frac{E}{L}$ and $q = \frac{1}{2} - \frac{E}{L}$, and where $E$ is a positive parameter. Note that, while our choices force $p > q$, we could just as easily address $p < q$ by swapping them. 

We proceed to examine the time evolution of the density for a site $i$, which results from adjacent particles hopping to $i$ and particles at $i$ hopping away
\begin{linenomath*}\begin{equation*}
\frac{d}{d\tau}\rho_\tau (i) = \frac{d}{d\tau}\mathds{E}^{\rho} \omega_\tau (i) = \mathds{E}^{\rho} p f(\omega_\tau (i-1)) + \mathds{E}^{\rho} q f(\omega_\tau (i+1)) - \mathds{E}^{\rho} p f(\omega_\tau (i)) - \mathds{E}^{\rho} q f(\omega_\tau (i)).
\end{equation*}\end{linenomath*} We now substitute the weak asymmetry condition in the following way
 \begin{linenomath*}\begin{align*}
	\frac{d}{d\tau} \mathds{E}^\rho \omega_\tau (i) &= -\mathds{E}^\rho f\,(\omega_\tau (i)) + \frac{1}{2} \mathds{E}^\rho f\,(\omega_\tau (i+1))\\ &- \frac{E}{L} \mathds{E}^\rho f\,(\omega_\tau (i+1)) + \frac{1}{2} \mathds{E}^\rho f\,(\omega_\tau (i-1)) + \frac{E}{L} \mathds{E}^\rho f\,(\omega_\tau (i-1))\\
	&= \frac{1}{2} \Big[ \mathds{E}^\rho f\,(\omega_\tau (i+1)) - 2 \mathds{E}^\rho f\,(\omega_\tau (i)) + \mathds{E}^\rho f\,(\omega_\tau (i-1)) \Big]\\ &- \frac{E}{L} \Big[ \mathds{E}^\rho f\,(\omega_\tau (i+1)) - \mathds{E}^\rho f\,(\omega_\tau (i-1)) \Big].
\end{align*}\end{linenomath*} We continue by defining $G(\rho) := \mathds{E}^\rho f\,(\omega)$ and substitute the rescaled $t$ and $x$ variables
\begin{linenomath*}\begin{align*}
	\frac{1}{dL^2} \frac{\partial}{\partial t} \mathds{E}^{\rho} \omega_{tdL^2} (xL) &= \frac{1}{2} \Big[ G\big(\rho_{tdL^2} (xL+1)\big) - 2 G\big(\rho_{tdL^2} (xL) \big) + G\big( \rho_{tdL^2} (xL -1) \big) \Big] \\
	& \quad - \frac{E}{L} \Big[ G\big( \rho_{tdL^2} (xL+1) \big) - G\big( \rho_{tdL^2} (xL -1) \big) \Big].
\end{align*}\end{linenomath*} Rearranging and identifying $\varrho_t (x)$, we find
\begin{linenomath*}\begin{align*}
\frac{\partial}{\partial t} \varrho_t (x) &= \frac{dL^2}{2} \Big[ G\left(\varrho_t \left(x + L^{-1}\right)\right) - 2 G\left(\varrho_t \left(x\right) \right) + G\left( \varrho_{t} \left(x - L^{-1}\right) \right) \Big] \\
	& \quad - d E L \Big[ G\left( \varrho_t \left(x+L^{-1}\right) \right) - G\left( \varrho_t \left(x - L^{-1}\right) \right) \Big].\\
&\simeq \frac{d}{2} \frac{\partial^2}{\partial x^2} G(\varrho_t(x)) - 2dE \frac{\partial}{\partial x} G (\varrho_t (x)).
\end{align*}\end{linenomath*}
We conclude
\begin{linenomath*}\begin{equation}\label{eq:hydro}
	\frac{\partial}{\partial t} \varrho_t(x) \simeq \frac{d}{2} \frac{\partial^2}{\partial x^2} G\big(\varrho_t (x) \big) - 2dE \frac{\partial}{\partial x} G\big(\varrho_t (x)\big).
\end{equation} \end{linenomath*}
To find the proper boundary conditions, we repeat the argument for the leftmost site
\begin{linenomath*}\begin{align*}
\frac{\partial}{\partial \tau} \mathds{E}^\rho \omega_\tau (1) &= \frac{1}{2} \left[ \mathds{E}^\rho f\,(\omega_\tau (2)) - \mathds{E}^\rho f\,(\omega_\tau (1)) \right] - \frac{E}{L} \left[ \mathds{E}^\rho f\,(\omega_\tau (2)) + \mathds{E}^\rho f\,(\omega_\tau (1)) \right]\\
\implies \frac{1}{L}\frac{\partial}{\partial t} \varrho_t (L^{-1}) &= \frac{dL}{2} \left[ G \left( \varrho_t \left(2L^{-1}\right) \right) - G \left( \varrho_t \left(L^{-1}\right) \right) \right] - dE \left[ G \left( \varrho_t \left(2L^{-1}\right) \right) + G \left( \varrho_t \left(L^{-1}\right) \right) \right]\\
&\simeq \frac{d}{2} \frac{\partial}{\partial x} G\left( \varrho_t (0) \right) - 2dE G\left( \varrho_t(0)\right).
\end{align*}\end{linenomath*} In the limit as $L\rightarrow\infty$, the $\frac{\partial}{\partial t}$ term drops out and we have the Robin boundary \begin{linenomath*}\begin{equation}\label{eq:lbc}
\frac{\partial}{\partial x} G \left( \varrho_t (0) \right) = 4 E G \left( \varrho_t (0) \right).
\end{equation}\end{linenomath*} Similarly, we obtain the following boundary condition for the rightmost site
%\begin{align*}
%\frac{\partial}{\partial \tau} \mathds{E}^\rho \omega_\tau (L) &= \frac{1}{2} \left[ \mathds{E}^\rho f\,(\omega_\tau (L-1)) - \mathds{E}^\rho f\,(\omega_\tau (L)) \right] + \frac{E}{L} \left[ \mathds{E}^\rho f\,(\omega_\tau (L-1)) + \mathds{E}^\rho f\,(\omega_\tau (L)) \right]\\
%\implies \frac{1}{L}\frac{\partial}{\partial t} \varrho_t (1) &= \frac{dL}{2} \left[ G \left( \varrho_t \left(1 - L^{-1}\right) \right) - G \left( \varrho_t \left(1\right) \right) \right] + dE \left[ G \left( \varrho_t \left(1-L^{-1}\right) \right) + G \left( \varrho_t \left(1\right) \right) \right]\\
%&\simeq -\frac{d}{2} \frac{\partial}{\partial x} G( \varrho_t (1)) + 2dE G( \varrho_t(1)).
%\end{align*} In the limit $L\rightarrow\infty$, 
\begin{linenomath*}\begin{equation}\label{eq:rbc}
\frac{\partial}{\partial x} G(\varrho_t (1)) = 4 E G (\varrho_t (1)).
\end{equation}\end{linenomath*} The boundary conditions \eqref{eq:lbc} and \eqref{eq:rbc} are consistent with the time-stationary solution of \eqref{eq:hydro}, together implying \begin{linenomath*}\begin{equation}\label{eq:statsol} \frac{\partial}{\partial x} G(\varrho_t (x)) = 4EG(\varrho_t (x)) \quad 0 \leq x \leq 1, \end{equation}\end{linenomath*} the general solution of which is $G (\varrho_t (x) ) = C e^{4Ex}$.

Equation~\ref{eq:hydro}, along with \eqref{eq:lbc} and \eqref{eq:rbc}, is the continuum description of the particle-based hillslope model. Note that, as in Section~\ref{sec:stationary}, this equation describes the evolution of the gradient process, and so its solutions must be integrated to obtain the corresponding hillslope profiles. In the special case of $f(\omega (i)) = \omega (i)$, $G(\varrho_t(x)) = \varrho_t (x)$, so the continuum equation is an advection-diffusion equation \begin{linenomath*}\begin{equation}\label{eq:specialhydro}
\frac{\partial}{\partial t} \varrho_t(x) \simeq \frac{d}{2} \frac{\partial^2}{\partial x^2} \varrho_t(x) - 2dE\frac{\partial}{\partial x} \varrho_t(x)
\end{equation}\end{linenomath*} with Robin boundary conditions \begin{linenomath*}\[ \frac{\partial}{\partial x} \varrho_t (0) = 4 E \varrho_t (0) \quad\text{and}\quad \frac{\partial}{\partial x} \varrho_t (1) = 4 E \varrho_t (1).\]\end{linenomath*} In the special case of $f(\omega (i)) = 1$ for $\omega (i) > 0$, $G(\varrho_t (x) ) = \varrho_t (x) / (1+\varrho_t (x))$, so the continuum equation has the following form \begin{linenomath*}\[ \frac{\partial}{\partial t} \varrho_t(x) \simeq \frac{d}{2} \frac{\partial^2}{\partial x^2} \frac{\varrho_t(x)}{1+\varrho_t (x)} - 2dE\frac{\partial}{\partial x} \frac{\varrho_t(x)}{1+\varrho_t (x)} \]\end{linenomath*} with Robin boundary conditions \begin{linenomath*}\[ \frac{\partial}{\partial x} \frac{\varrho_t (0)}{1+\varrho_t (0)} = 4 E \frac{\varrho_t (0)}{1+\varrho_t (0)} \quad\text{and}\quad \frac{\partial}{\partial x} \frac{\varrho_t (1)}{1+\varrho_t (1)} = 4 E \frac{\varrho_t (1)}{1+\varrho_t (1)}.\]\end{linenomath*} Appendix~\ref{sec:fokker} describes the solution to \eqref{eq:hydro} subject to the boundary conditions \eqref{eq:lbc} and \eqref{eq:rbc}.

\subsection{Scaling recap}\label{sec:scalingrecap}
We recall some key points from Section~\ref{sec:hydro} before describing simulations and dimensionalization.\begin{enumerate}
\item The scaling procedure consists of three steps: balancing incoming and outgoing particles, substituting the weak asymmetry condition, and substituting the rescaled variables.
\item The resulting continuum equation describes the density of gradient particles and is of advection-diffusion type.
\item The continuum equation contains a function $G$ which has simple, explicit forms when the rate function is linear or constant.
\item The scaling argument confirms that, even if the continuum equation is complicated, its solutions can easily be approximated by simulating the corresponding particle model.
\end{enumerate}

\section{Simulation and dimensionalization}\label{sec:sim}

The analysis of Section~\ref{sec:hydro} tells us that if we want to study the evolution of hillslopes according to \eqref{eq:hydro}, we can simulate the particle model of Section~\ref{sec:model} instead. As choices of rate $f(\omega) \neq \omega$ generally lead to a nonlinear PDE \eqref{eq:hydro}, simulating the particle model may often be preferable to an analytic approach or a numerical scheme. In addition to simulating the equilibrium hillslope profiles under various choices of $p$ and rate function $f$, we would also like to simulate the response of hillslopes to perturbations, such as river erosion or climate change (usually implemented by a change in a diffusion coefficient \citep{fernandes1997hillslope,mudd2004influence,roering2001hillslope}). We begin with simulations of equilibrium hillslope profiles.

\subsection{Equilibrium hillslope profiles}\label{sec:simequil}

\begin{figure}[htbp]
%\centering
\begin{tikzpicture}
    \node[anchor=south west,inner sep=0] (image) at (0,0) {\includegraphics[width=\textwidth]{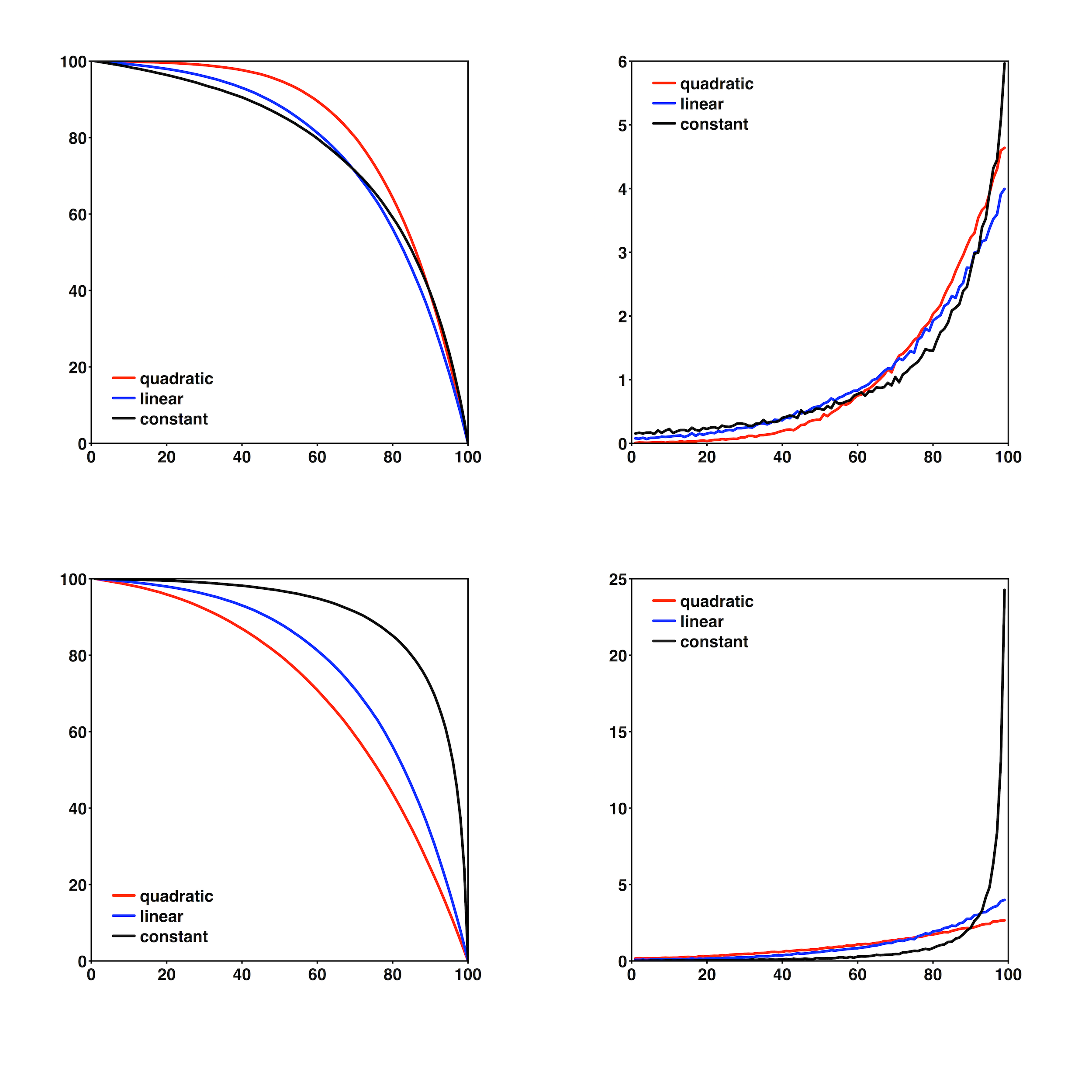}};
    \begin{scope}[x={(image.south east)},y={(image.north west)}]
        \node at (0.02,0.77) {\LARGE{\textbf{$h (i)$}}};
        \node at (0.02,0.3) {\LARGE{\textbf{$h (i)$}}};
        \node at (0.515,0.77) {\LARGE{\textbf{$\omega (i)$}}};
        \node at (0.515,0.3) {\LARGE{\textbf{$\omega (i)$}}};
        \node at (0.255,0.55) {{\LARGE{\textbf{$i$}}}};
        \node at (0.255,0.07) {{\LARGE{\textbf{$i$}}}};
        \node at (0.75,0.55) {{\LARGE{\textbf{$i$}}}};
        \node at (0.75,0.07) {{\LARGE{\textbf{$i$}}}};
        
        \node at (0.02,0.95) {\huge{\textbf{A}}};
        \node at (0.52,0.95) {\huge{\textbf{B}}};
        \node at (0.02,0.48) {\huge{\textbf{C}}};
        \node at (0.52,0.48) {\huge{\textbf{D}}};
        
        %\draw[help lines,xstep=.1,ystep=.1] (0,0) grid (1,1);
        %\foreach \x in {0,1,...,9} { \node [anchor=north] at (\x/10,0) {0.\x}; }
        %\foreach \y in {0,1,...,9} { \node [anchor=east] at (0,\y/10) {0.\y}; }
    \end{scope}
\end{tikzpicture}
\caption[Choice of rate changes profile]{Equilibrium hillslope (\textbf{A} and \textbf{C}) and gradient profiles (\textbf{B} and \textbf{D}) for quadratic \big($f(\omega) = {\omega}^2$\big), linear \big($f(\omega) = \omega $\big), and constant \big($f(\omega) = 1$ if $\omega > 0$ \big) rates. For \textbf{A} and \textbf{B}, parameters were $p=0.52$ (quadratic), $p=0.51$ (linear), $p=0.505$ (constant), $H=L=100$. For \textbf{C} and \textbf{D}, parameters were $p=0.51$ (all rates), $H=L=100$. All curves were obtained as the average over 10 identical trials.}
\label{fig:rate_profile}
\end{figure}
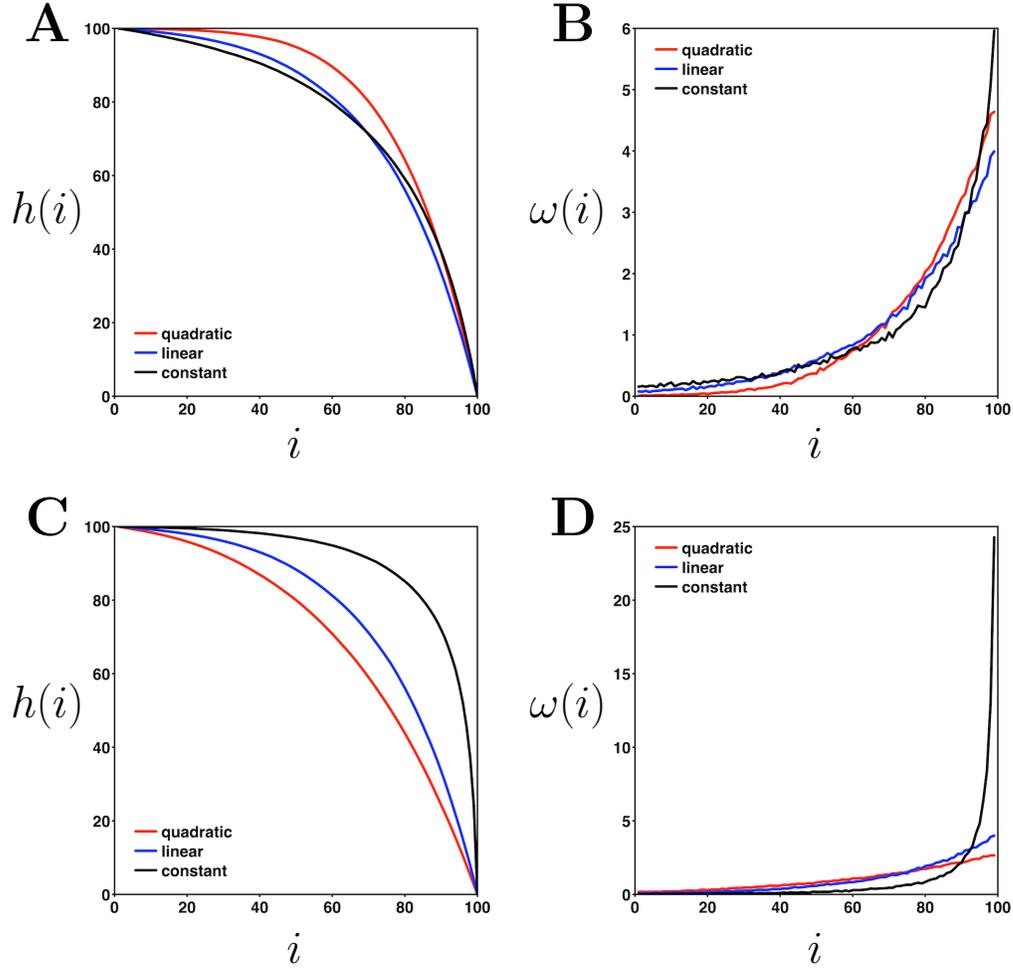

When the hopping rates of the gradient particle system are chosen to be $f(\omega (i)) =  \omega (i)$, the hillslope gradients satisfy Equation~\ref{eq:specialhydro}, which is solved by a drifting diffusion. For other choices of rates, the gradients evolve according to Equation~\ref{eq:hydro}. \citet{balazs2007convexity} showed that convex (concave) $f(\omega (i))$ implies convexity (concavity) of $G(\rho)$. To demonstrate these two cases, we pick constant and quadratic rates given by \big($f(\omega (i)) = 1$ for $\omega (i) > 0$, $f(\omega (i)) = 0$ for $\omega (i) = 0$ \big) and $f(\omega (i)) = {\omega (i)}^2$, respectively. As a result of Section~\ref{sec:hydro}, the behavior of these solutions can be understood by simulating the corresponding particle model. Stationary hillslope and gradient profiles are compared in Figure~\ref{fig:rate_profile}. In particular, Figure~\ref{fig:rate_profile}A and \ref{fig:rate_profile}B highlight that, for different choices of $p$, the profiles arising from linear, quadratic, and constant rates can be made relatively similar, but their curvatures differ. Figure~\ref{fig:rate_profile}C and \ref{fig:rate_profile}D show that, when $p$ is fixed, the profile arising from a constant rate is far steeper than those from linear and quadratic rates. Note that the profiles in the linear rate case can be calculated from (\eqref{eq:heights}), while the constant and quadratic results can be produced with the following simulation procedure.

We begin by specifying $f(\omega (i))$, parameters $H$, $L$, $p$, and the number of simulation time steps, $N$. We choose an initial height profile, which satisfies the boundary conditions, and use $\omega (i) = h (i) - h (i+1)$ to get the corresponding gradient profile. For each time step, we (i) apply $f(\omega (i))$ to $\omega (i)$, (ii) sample hop latencies from independent exponential distributions with rates $f(\omega (i))$, and (iii) update $\omega (i)$ and $h (i)$ to reflect the first hop, contingent on satisfying boundary conditions. We implemented this procedure and conducted all simulations in MATLAB (R2016b, The MathWorks, Inc., Natick, Massachusetts, United States); the code can be obtained by emailing the corresponding author.

\subsection{Hillslope perturbations and empirical flux}\label{sec:simperturb}

\begin{figure}[htbp]
\centering
\begin{tikzpicture}
    \node[anchor=south west,inner sep=0] (image) at (0,0) {\includegraphics[height=0.8\textheight]{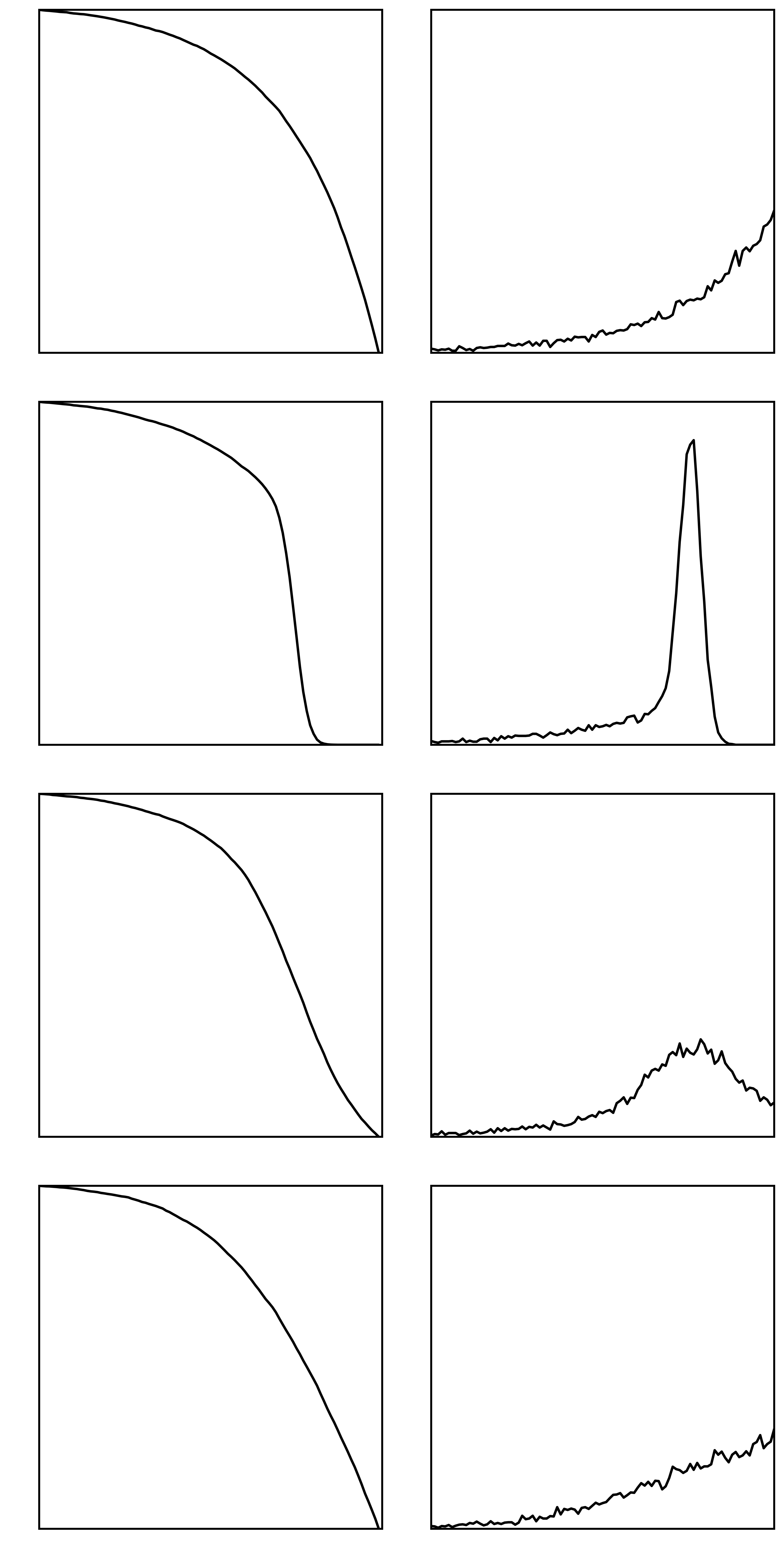}};
    \begin{scope}[x={(image.south east)},y={(image.north west)}]
        \node at (-0.01,0.985) {\Large{\textbf{$H$}}};
	\node at (-0.01,0.885) {\Large{\textbf{$h(i)$}}};
        \node at (-0.01,0.785) {\Large{\textbf{$0$}}};
        
        \node at (1.048,0.985) {\Large{\textbf{$\rho (L)$}}};
        \node at (1.048,0.885) {\Large{\textbf{$\omega (i)$}}};
        \node at (1.035,0.785) {\Large{\textbf{$0$}}};       
         
        \node at (0.437,0) {\Large{\textbf{$L+1$}}};
        \node at (0.05,0) {\Large{\textbf{$1$}}};
        
        \node at (0.967,0) {\Large{\textbf{$L$}}};
        \node at (0.55,0) {\Large{\textbf{$1$}}};
        %\node at (0.55,0) {\Large{\textbf{$0$}}};
        
%        \node at (-0.12,0.885) {\huge{\textbf{A}}};
%        \node at (-0.12,0.635) {\huge{\textbf{B}}};
%        \node at (-0.12,0.385) {\huge{\textbf{C}}};
%        \node at (-0.12,0.135) {\huge{\textbf{D}}};

        \node at (0.1,0.80) {\huge{\textbf{A}}};
        \node at (0.1,0.55) {\huge{\textbf{B}}};
        \node at (0.1,0.30) {\huge{\textbf{C}}};
        \node at (0.1,0.05) {\huge{\textbf{D}}};
        
        \node at (0.265,1.015) {\Large{\textbf{Height}}};
        \node at (0.765,1.015) {\Large{\textbf{Gradient}}};
        
        %\node at (-0.01,0.615) {\huge{\textbf{$t$}}};
        %\node[rotate=-90] at (-0.005,0.50) {\Large{\textbf{$\xrightarrow{\makebox[3cm]{}}$}}};
        
        %\draw[help lines,xstep=.1,ystep=.1] (0,0) grid (1,1);
        %\foreach \x in {0,1,...,9} { \node [anchor=north] at (\x/10,0) {0.\x}; }
        %\foreach \y in {0,1,...,9} { \node [anchor=east] at (0,\y/10) {0.\y}; }
    \end{scope}
\end{tikzpicture}
\caption[Response to perturbation.]{Simulated hillslope response to a river-erosion-like perturbation. A hillslope in equilibrium (\textbf{A}) with linear rate $f(\omega)$ = $\omega$ is perturbed (\textbf{B}) and relaxes (\textbf{C} and \textbf{D}). The rows depict time steps $1$ $\times$ $10^6$ (\textbf{A}), $1.1 \times 10^6$ (\textbf{B}), $2.5 \times 10^6$ (\textbf{C}), and $5 \times 10^6$ (\textbf{D}), for a perturbation applied near the righthand boundary immediately after time step $1\times 10^6$. The particle system was initialized at equilibrium (Equation~\ref{eq:heights}) with parameters $p=0.51$, $H=1\times 10^4$, and $L=100$. At equilibrium, $\rho (i)$ is given by Equation~\ref{eq:density}.}
\label{fig:vert_panels}
\end{figure}
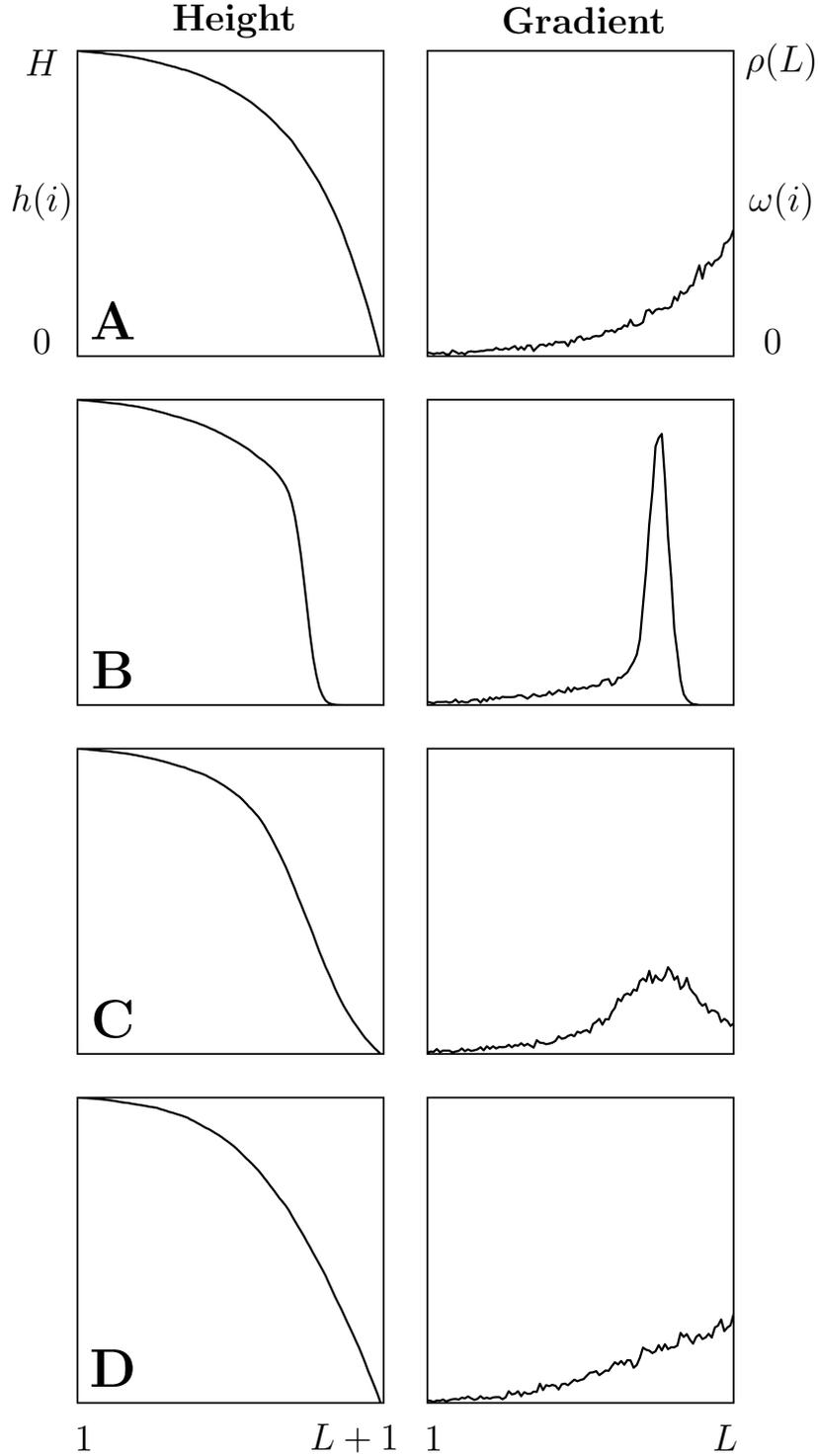

We now turn our attention to hillslopes perturbed away from equilibrium, to study the timescales over which hillslopes relax and the influence the parameters have over this process. Consider the gradient process with $f(\omega (i)) = \omega (i)$, $L = 100$, $H=1\times 10^4$, and $p = 0.51$. We initialize the process with $\omega (i)$ corresponding to $\texttt{ceil}(h (i))$, where the $h (i)$ are given by Equation~\ref{eq:heights}. We introduce a river-erosion-inspired perturbation, which conserves total gradient particle count, by skimming $50$ gradient particles from each site with at least that many. All of the skimmed particles are added to a single site, and we track $h (i)$ and $\omega (i)$ as the hillslope relaxes back to equilibrium (Figure~\ref{fig:vert_panels}). Figure~\ref{fig:vert_panels}A depicts the hillslope and gradient profiles maintaining equilibrium after $1$ $\times$ $10^6$ timesteps. Immediately after this frame, the perturbation was applied. Figure~\ref{fig:vert_panels}B and \ref{fig:vert_panels}C show the profiles smoothing and refilling the base at timesteps $1.1$ $\times$ $10^6$ and $2.5$ $\times$ $10^6$, respectively. By timestep $5$ $\times$ $10^6$, the hillslope resembles the equilibrium hillslope. 

\begin{figure}[htbp]
\centering
\begin{tikzpicture}
    \node[anchor=south west,inner sep=0] (image) at (0,0) {\includegraphics[width=0.8\textwidth]{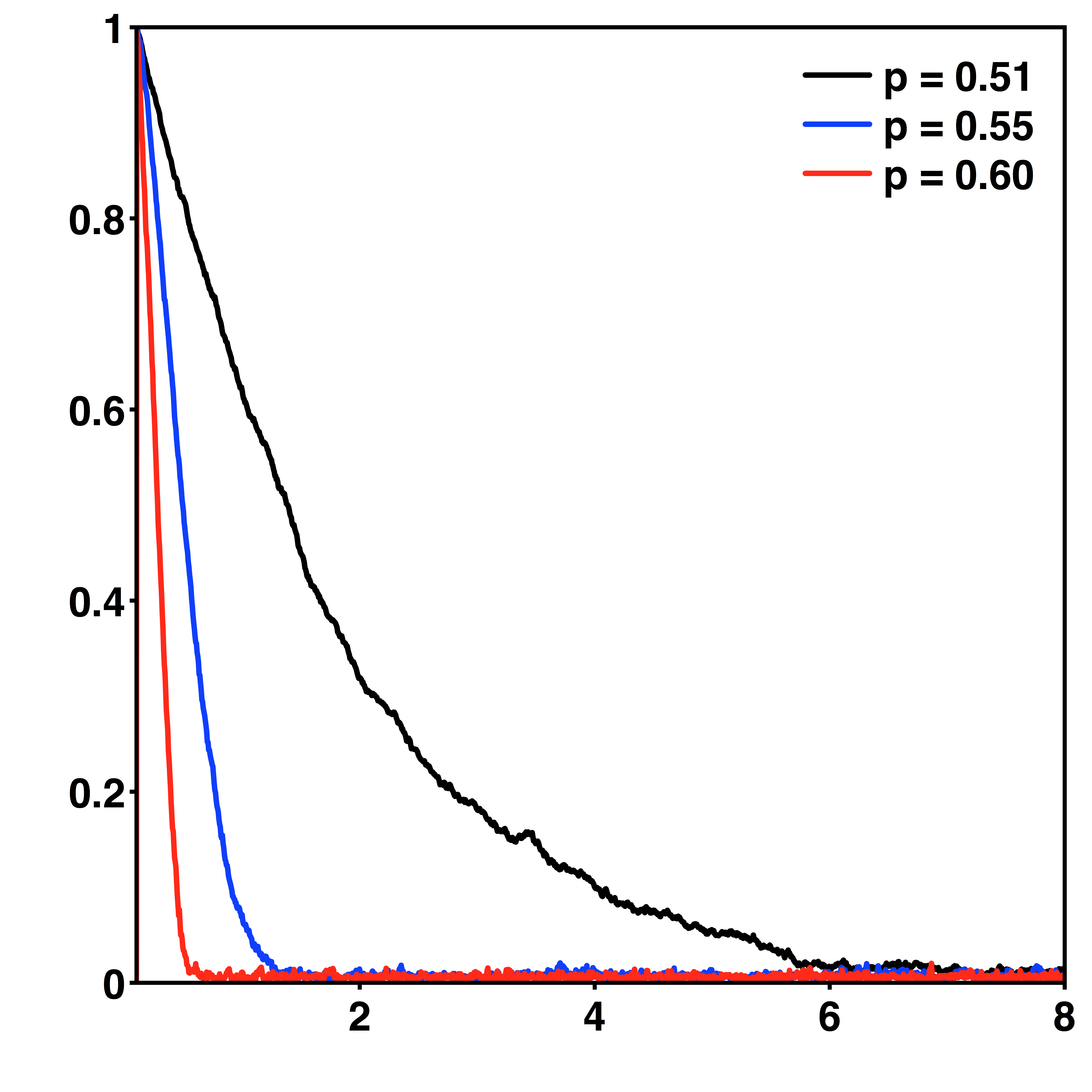}};
    \begin{scope}[x={(image.south east)},y={(image.north west)}]
        \node at (0,0.54) {\huge{\textbf{$\Delta h_t $}}};
        \node at (0.55,0.02) {{\huge{\textbf{$t$}}} {\Large($\times \, 10^7$ steps)}};
        %\draw[help lines,xstep=.1,ystep=.1] (0,0) grid (1,1);
        %\foreach \x in {0,1,...,9} { \node [anchor=north] at (\x/10,0) {0.\x}; }
        %\foreach \y in {0,1,...,9} { \node [anchor=east] at (0,\y/10) {0.\y}; }
    \end{scope}
\end{tikzpicture}
\caption[Relaxation to equilibrium.]{Hillslope profile relaxation in response to a perturbation, for a particle system with $f(\omega (i)) = \omega (i)$, $p=0.51$, $p=0.55$, or $p=0.60$, $H=1\times 10^4$, $L=100$, and $t = 0$ to $t = 8 \times 10^7$. $\Delta h_t$ (defined by \eqref{eq:relaxdist}) was normalized by its largest value over the simulation. Each curve is the average over 25 trials.}
\label{fig:decay}
\end{figure}

It is natural to wonder about the affect $p$ has on the rate of hillslope relaxation in response to perturbations which do not change the underlying dynamics. Consider the same process, with $p = 0.51$, $p = 0.55$, or $p = 0.60$. Take \begin{linenomath*}\begin{equation}\label{eq:relaxdist}\Delta h_t (i) : = \big| h_t (i) - h_0 (i) \big| \qquad\text{and}\qquad \Delta h_t := \sum_{i=1}^{L} \Delta h_t (i) \end{equation}\end{linenomath*} as measures of distance from the $h_0$ equilibrium. The results for $t = 0$ to $t = 8 \times 10^7$ are shown in Figure~\ref{fig:decay}. It seems that the larger $p$ is, the greater the asymmetry in hopping rates, and the faster the hillslope returns to equilibrium. However, the perturbation depends on the gradient profile, and larger values of $p$ are associated with steeper hillslopes, meaning the local slope is not controlled in the experiment.

\begin{figure}[htbp]
%\centering
\begin{tikzpicture}
    \node[anchor=south west,inner sep=0] (image) at (0,0) {\includegraphics[width=\textwidth]{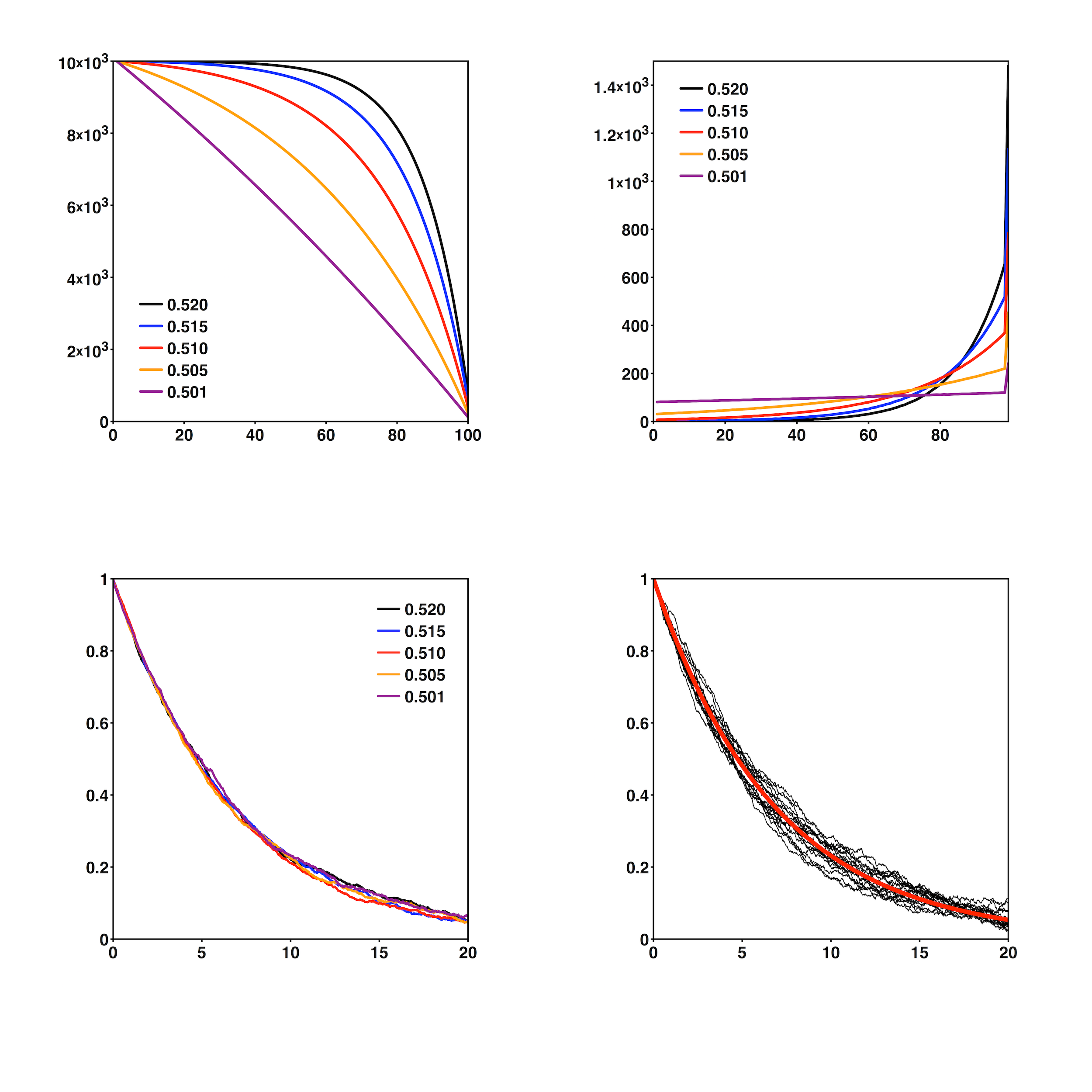}};
    \begin{scope}[x={(image.south east)},y={(image.north west)}]
        \node at (0.02,0.77) {\LARGE{\textbf{$h (i)$}}};
        \node at (0.02,0.3) {\LARGE{\textbf{$\Delta h_t (i)$}}};
        \node at (0.515,0.77) {\LARGE{\textbf{$\omega (i)$}}};
        \node at (0.515,0.3) {\LARGE{\textbf{$\Delta h_t (i)$}}};
        \node at (0.27,0.55) {{\LARGE{\textbf{$i$}}}};
        \node at (0.27,0.07) {{\LARGE{\textbf{$t$}}} {\Large($\times \, 10^3$ steps)}};
        \node at (0.76,0.55) {{\LARGE{\textbf{$i$}}}};
        \node at (0.76,0.07) {{\LARGE{\textbf{$t$}}} {\Large($\times \, 10^3$ steps)}};
        
        \node at (0.015,0.95) {\huge{\textbf{A}}};
        \node at (0.515,0.95) {\huge{\textbf{B}}};
        \node at (0.015,0.48) {\huge{\textbf{C}}};
        \node at (0.515,0.48) {\huge{\textbf{D}}};
        
        %\draw[help lines,xstep=.1,ystep=.1] (0,0) grid (1,1);
        %\foreach \x in {0,1,...,9} { \node [anchor=north] at (\x/10,0) {0.\x}; }
        %\foreach \y in {0,1,...,9} { \node [anchor=east] at (0,\y/10) {0.\y}; }
    \end{scope}
\end{tikzpicture}
\caption[The role of $p$ and gradient on hillslope relaxation in the linear rate model]{The role of $p$ and gradient on hillslope relaxation in the linear rate model. Equilibrium hillslope profiles for a variety of choices of $p$ (\textbf{A}) and the corresponding gradient profiles (\textbf{B}). Parameters were $H$ = $1$ $\times $ $10^4$ and $L$ = $100$, with linear rate $f (\omega)$ = $\omega$. The gradient profiles overlap around $i$ = $70$, so we can control for the affect of local slope on the rate of hillslope relaxation by perturbing in the overlap region. For each choice of $p$, the perturbation (applied at the beginning of the simulation) consisted of taking one quarter of the gradient particles from each of 10 sites in an interval centered on $i$ = $70$, and adding them all to the leftmost site in the interval. The resulting time series of $\Delta h_t (i)$ (defined by \eqref{eq:relaxdist}) were well-fit by exponential decay with common time constant $1.47\times 10^{-4}$ (\textbf{C}). $R^2$ $>$ $0.995$ in all cases. In (\textbf{D}), we fixed $p$ = $0.51$ and performed the perturbation experiment using a sliding, 10-site window, in order to test various local gradients along the hillslope. The resulting, normalized $\Delta h_t $ decays for $i = 10, 15, \dots, 85, 90$ are shown (thin black curves) with the exponential fit superimposed (thick red curve). Each curve in \textbf{C} and \textbf{D} was the average of 25 identical trials.}
\label{fig:independent_response_profiles}
\end{figure}

To separately test the affects of $p$ and local slope on the rate of hillslope relaxation, we identified contiguous, 10-site regions of equilibrium hillslopes, for various choices of $p$, which had slope similar to that of an equilibrium profile for a different choice of $p$ (Figure~\ref{fig:independent_response_profiles}A and \ref{fig:independent_response_profiles}B). We then perturbed these regions of similar slope by adding one quarter of the total number of gradient particles in that region to a single drop site. For the linear rate model, the time series of $\Delta h_t (i)$ (where $i$ was the drop site) were well-fit by exponential decays ($R^2 > 0.995$ in all cases) with identical time constants (Figure~\ref{fig:independent_response_profiles}C). We then conducted the same perturbation, but for all possible contiguous 10-site windows. The resulting exponential decays for sites $i=10, 15, \dots, 90$ had time constants which agreed with that of Figure~\ref{fig:independent_response_profiles} and are summarized in (Figure~\ref{fig:independent_response_profiles}D). These simulation results suggest that, for linear rate, the timescale over which hillslopes relax does not depend on $p$ or the local slope; this conclusion is in agreement with the calculation of Section~\ref{sec:fokker} and \eqref{eq:relax} in particular. We emphasize that this is \textit{not} the case in general.

Fluxes develop along the hillslope during the process of equilibration which, while not directly accessible via the methods of Section~\ref{sec:hydro}, can be approximated by an ``empirical'' flux inferred from height changes along the hillslope. For example, growth downslope of site $i$ suggests that a flux arose upslope of site $i$. As this indirect measurement of flux relies on height changes, it depends on two observations times $t$ and $t + \Delta t$. We calculate the empirical flux at site $i$, relative to time steps $t$ and $t + \Delta t$ as
\begin{linenomath*}\begin{equation}\label{eq:empiricalflux}
\phi_{t+\Delta t} (i) - \phi_t (i) = r \, \Delta t + \sum_{j > i} \big( h_{t+\Delta t} (j) - h_t (j) \big).
\end{equation}\end{linenomath*}
Here, $r$ is a constant flux coming from the right boundary and we adopt the convention that a positive value of flux at a site $i$ indicates a net, relative height change for sites $j > i$. %In Appendix~\ref{app:flux}, we provide a heuristic argument to identify a corresponding continuum flux. 

\begin{figure}[htbp]
\centering
\begin{tikzpicture}
    \node[anchor=south west,inner sep=0] (image) at (0,0) {\includegraphics[width=\textwidth]{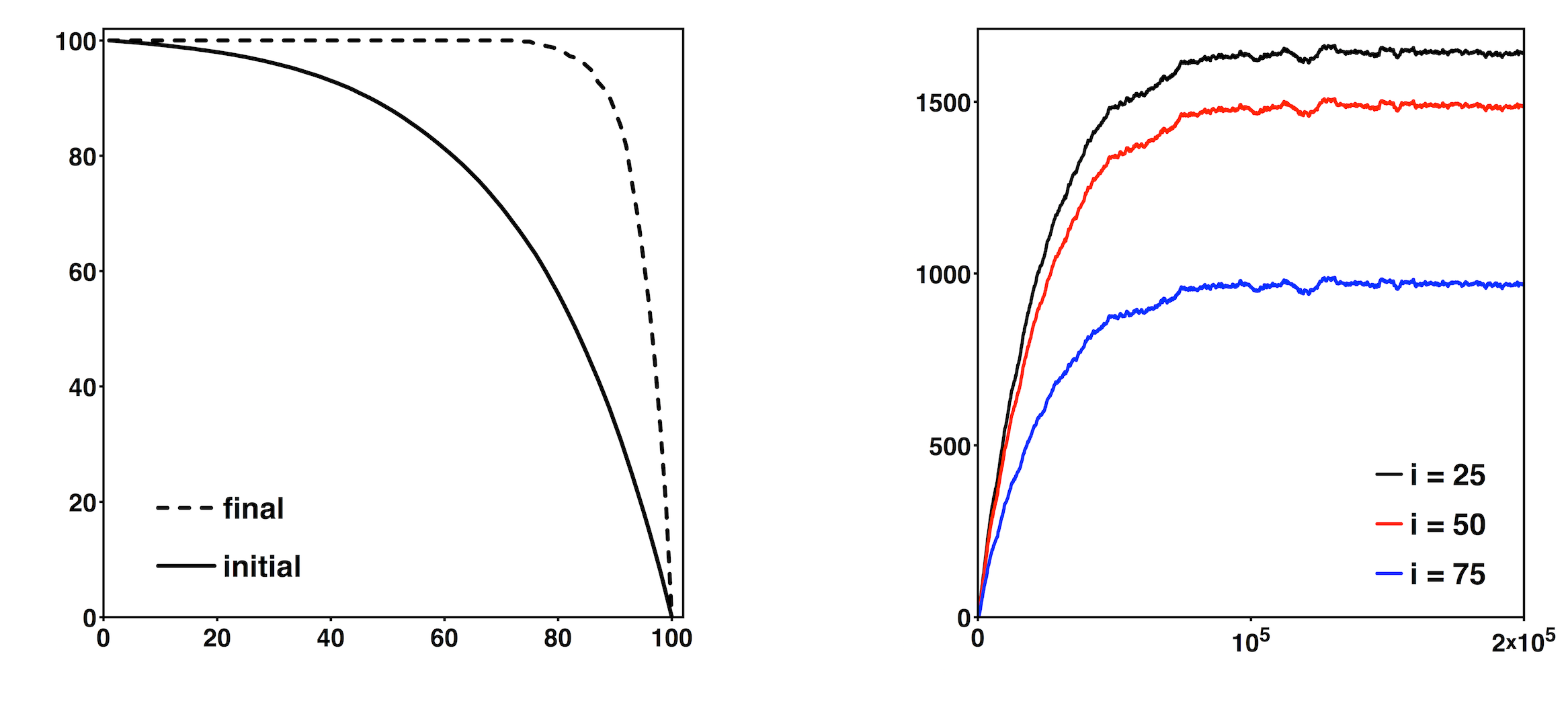}};
    \begin{scope}[x={(image.south east)},y={(image.north west)}]
        \node at (0.01,0.54) {\LARGE{\textbf{$h(i)$}}};
        \node at (0.25,0.04) {\LARGE{\textbf{$i$}}};
        \node at (0.55,0.54) {\LARGE{\textbf{$\overline{\phi}_t (i)$}}};
        \node at (0.8,0.04) {\LARGE{\textbf{$t$}}};
        
        \node at (0.01,0.92) {\huge{\textbf{A}}};
        \node at (0.55,0.92) {\huge{\textbf{B}}};
        %\draw[help lines,xstep=.1,ystep=.1] (0,0) grid (1,1);
        %\foreach \x in {0,1,...,9} { \node [anchor=north] at (\x/10,0) {0.\x}; }
        %\foreach \y in {0,1,...,9} { \node [anchor=east] at (0,\y/10) {0.\y}; }
    \end{scope}
\end{tikzpicture}
\caption[Empirical fluxes during equilibration.]{A hillslope equilibrated for $p$ = $0.51$, $H$ = $100$, $L$ = $100$, and linear rate $f(\omega)$ = $\omega$, is perturbed by an abrupt change in the dynamics to $p$ = $0.55$. In \textbf{A}, the initial profile (solid line) evolves with updated $p$ to the final, near-equilibrium hillslope (dotted line). In \textbf{B}, cumulative fluxes $\overline{\phi}_t (i)$ (defined by \eqref{eq:phibar}) develop in response to the perturbation. Cumulative fluxes are shown for sites $i=25$, $i=50$, and $i=75$, averaged over 100 identical trials. By convention, a flux at site $i$ is positive if it indicates net hillslope height increase for sites $j > i$.}
\label{fig:flux}
\end{figure}
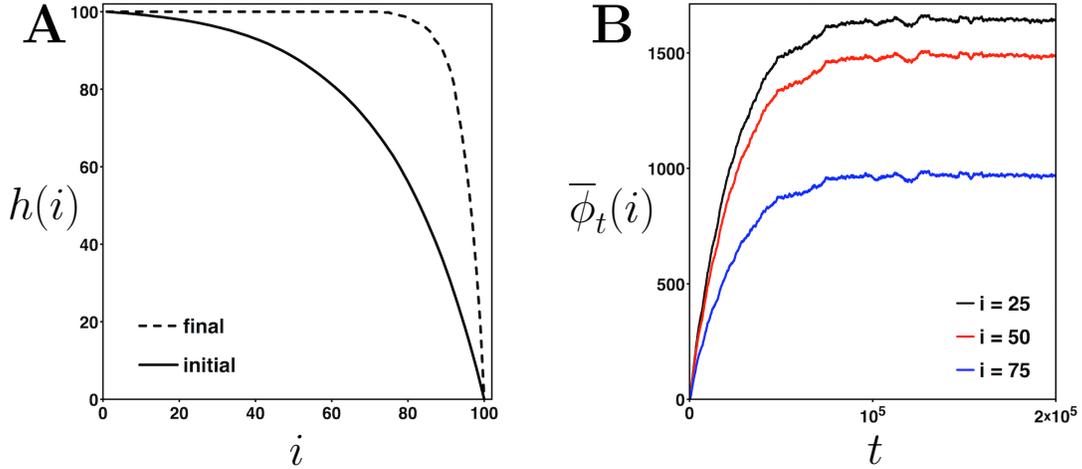

To demonstrate the use of the empirical flux, we consider a hillslope with $H=L=100$, initially at equilibrium with $p=0.51$. For convenience, we choose $\Delta t$ to be the length of one time step in the simulation. Immediately after $t=0$, we switch to $p=0.55$, producing a net positive flux toward the righthand boundary, as the hillslope tries to equilibrate. To isolate the flux contributions driven by equilibration from those of the constant flux $r$, we instead track the \textit{cumulative} flux through site $i$ as
\begin{linenomath*}\begin{equation}\label{eq:phibar} 
\overline{\phi}_{t} (i) := \phi_{t} (i) - \phi_{0} (i) - r t.
\end{equation}\end{linenomath*} Figure~\ref{fig:flux}A shows the before-and-after hillslope profiles, corresponding to $p=0.51$ and $p=0.55$, and Figure~\ref{fig:flux}B shows the cumulative flux through sites $i=25$, $50$, and $75$ during equilibration.
%\begin{equation}\label{eq:phibar} 
%\overline{\phi}_{t} (i) := \sum_{t' = 1}^t \left( \phi_{t'} (i) - \phi_{t'-1} (i)\right) - r t.
%\end{equation} 

\subsection{Adding dimensions and fitting parameters}\label{sec:dimension}

In order to reliably translate simulation results into empirically testable predictions, we need a principled way of assigning dimensions to otherwise dimension-less model quantities (e.g. particle model length $L$ and the length $\ell$ of an observed hillslope, in meters). Additionally, we need to specify how hillslope data are used to fit model parameters. We suggest the following procedure, which is partly motivated by the calculations in Appendix~\ref{sec:fokker}.

%Consider instead the reverse process of removing dimensions from a model of a particular geomorphic process. We would identify time- and length-scales which are, in some sense, characteristic of the process. We would then ``divide-out'' the dimensions of model parameters and observables, using these time- and length-scales (as in \cite{mudd2004influence} and \cite{shelef2016unified}). %We adopt the view that our dimensionless $L$ and $H$ are the result of ``dividing-out'' the dimensions of length from an observed hillslope by a characteristic length scale (as in \cite{mudd2004influence} and \cite{shelef2016unified}). To reverse this process, we can choose a characteristic length scale and multiply $L$ and $H$ by it. 

Recall that sites in the particle model of Section~\ref{sub:dynamics} are indexed by $i = 1,2,\dots,L$. Let $i$ count the number of sediment grains in the length of the hillslope. If a typical grain has a diameter of $2$ millimeters and the hillslope length is measured to be $\ell = 200$ meters, then set $L$ = $200$ meters / $2$ millimeters = $100\,000$. Similarly, if the crest of the hillslope is $h = 100$ meters above the height at the end of the hillslope (at a distance $\ell$ meters from the crest), assign $H$ = $100$ meters / $2$ millimeters = $50\,000$. In this way, we relate dimensionless particle model quantities $L$ and $H$ to observable hillslope quantities with dimension, $\ell$ and $h$. 

We now consider fitting $E$, which encapsulates the asymmetry in the underlying gradient process, and adding dimension to the simulation timesteps. For simplicity, we consider the case of the linear rate model, but the following procedure can be applied to nonlinear rate models using the contents of Appendix~\ref{app:nonlin}. We can estimate the parameter $E$ from measurements of the equilibrium or near-equilibrium shape of the hillslope, by fitting \eqref{eq:hx}. Next, we can add dimension to the simulation timesteps by fitting the time constant $d$, which was introduced in the scaling argument of Section~\ref{sec:hydro}. Fitting $d$ requires that a small perturbation $r_0$ be added to the hillslope, the relaxation of which obeys \eqref{eq:relax}. Ideally, the location of the perturbation and the timescale of relaxation should be such that the boundaries do not play a significant role. To summarize, we suggest the following, three-step approach. \begin{enumerate} \item Measure typical grain diameter to add units to $H$ and $L$. \item Fit $E$ to equilibrium hillslope shape. \item Fit $d$ to hillslope relaxation in response to a perturbation.\end{enumerate}

While the first step does not depend on the choice of rate function, the second and third steps do, as the form of the rate affects the the relationship between $E$ and the equilibrium hillslope shape, and relationship between $d$ and the relaxation of perturbations. We also note that this procedure makes use of both small-scale and large-scale measurements, as well as information about hillslope equilibrium and nonequilibrium.

\subsection{Simulation recap}\label{sec:simrecap}
We collect some key points from Section~\ref{sec:sim} before continuing on to the discussion.\begin{enumerate}
\item We simulated perturbations in two ways: rearranging the gradient particles (through $\omega$) and changing the dynamics (through $p$ or, equivalently, $E$).
\item Hillslope relaxation in response to perturbation can be tracked by comparing it with the corresponding stationary profile or by tracking the empirical fluxes.
\item In the linear rate case, hillslope relaxation timescale is independent of $E$, $H$, and $L$ .
\item The simulation results can be assigned dimensions to facilitate comparison with observations, according to the procedure of Section~\ref{sec:dimension}.
\end{enumerate}

\section{Discussion}\label{sec:disc}

The key ingredient of the particle model of Section~\ref{sec:model} is indirection: the decision for particles to represent units of hillslope gradient, instead of units of hillslope height. Consider again the scenario of Figure~\ref{fig:fixedheightzrp}. Had we specified similar dynamics on the hillslope profile directly, the resulting profiles could be unrealistic (e.g. large particle build-up next to sites with no particles) and the dynamics would require awkward constraints to prevent such profiles. Most importantly, this process would not have stationary profiles which are amenable to analysis, and a scaling argument like that of Section~\ref{sec:hydro} would not apply. In this sense, the gradient particle model is a natural choice, but one made at the expense of direct access to information about sediment flux and particle hopping distances. Indeed, although we can obtain the hillslope profile from the gradient particle profile (using \eqref{eq:inversion}), our model does not prescribe a dynamics on the hillslope profile and so is agnostic to fluxes of hillslope particles and the distances they typically travel. Critically, this circumvents the issue of specifying whether transport on the hillslope is local or nonlocal and, as a result, our model can represent a variety of geomorphic processes and the scaling argument holds across transport regimes. 

We are free to accessorize our model with fluxes, defined in terms of hillslope gradient, which evolve according to the particle model of Section~\ref{sec:model} or, in the continuum, according to \eqref{eq:hydro}. In Section~\ref{sec:simperturb}, for example, we proposed a nonlocal flux \eqref{eq:empiricalflux} in terms of changes in the hillslope height (equivalently, changes in hillslope gradient via \eqref{eq:inversion}). %and scaled it to a continuum flux equation \eqref{eq:finalflux} in Appendix~\ref{app:flux}. 
Alternatively, we could specify a local flux like those of \citet{culling1963soil} (linear dependence on slope), \citet{andrews1987fitting} (nonlinear dependence on slope), and \citet{furbish2009rain} (nonlinear, includes height and slope), or a nonlocal flux of the form favored by \citet{furbish2013sediment}. This freedom reflects the \textit{hillslope-first} nature of our particle model, for which we formulate the dynamics of the hilllslope gradients and infer the flux, as opposed to formulating the dynamics of the flux, from which we then infer the hillslope profile.

Such a hillslope-first approach may be more natural than a nonlocal, transport-first approach for conducting perturbation experiments like those described in Section~\ref{sec:simperturb}. For example, consider the experiment illustrated by Figures~\ref{fig:vert_panels} and \ref{fig:decay}, which simulates hillslope recovery from river erosion. Nonlocal formulations of transport require as input a distribution of particle travel distances \citep{furbish2010divots} or an assumption about the degree of nonlocality \citep{foufoula2010nonlocal}, but these features depend on the hillslope gradient, and so should vary throughout the experiment \citep{gabet2012particle}. In contrast, our model fixes the law governing the redistribution of hillslope gradient through the rate function $f$, which is an input of the modeler.

Given a choice of $f$, the parameter $p$ can be determined from an observation of hillslope shape, according to the procedure described in Section~\ref{sec:dimension}. Intuitively, for a given rate function, $p > \frac{1}{2}$ specifies a deposition-type process; $p < \frac{1}{2}$ specifies a washing-out-type process. For example, in Figure~\ref{fig:exp_curves}, $p = 0.49$ produces a stationary hillslope profile resembling one formed under sheet wash with gullies, while $p=0.51$ results in a profile which more closely resembles one formed under soil creep. The parameter $p$ can also be used to conduct perturbation experiments, as in Figure~\ref{fig:flux}, where the hillslope begins as the stationary profile under a process associated with $p=0.51$ and must equilibrate after an external driver (e.g. climate change) alters the dynamics to $p=0.55$. Unlike the case of a river-erosion-like perturbation, it may be possible to use a nonlocal, transport-first approach to conduct similar experiments, for example, by making a small change to a parameter in the distribution of particle travel distances.

The particle-based model of Section~\ref{sec:model} is purely probabilistic, unlike those of \citet{kirkby1975surface}, \citet{gabet2012particle}, which incorporate frictional forces associated with particle motion, and that of \citet{dibiase2017slope}, which also accounts for variations in grain size and is extended to motion in two spatial dimensions. These approaches benefit from directly incorporating hillslope microtopography, but are computationally-expensive in a way which may prohibit the simulation of hillslope evolution over long timescales \citep{dibiase2017slope}, and cannot be scaled to corresponding continuum equations \citep{ancey2015stochastic}. Our model is most similar to that of \citet{tucker2010trouble}, which is also purely probabilistic, rules-based, and computationally-inexpensive, but for which a corresponding continuum description is unavailable.

The scaling argument of Section~\ref{sec:hydro} claims that, under the appropriate scalings of time and space variables, and in the limit as $L\rightarrow\infty$, the model behaves according to an advection-diffusion equation. Note that this governs the scaled gradient process, not the hillslope itself -- we must integrate the solutions to obtain the corresponding hillslope. For the linear rate case, we can solve the continuum equation directly; in the nonlinear rate case, numerical methods may be required. Both the scaling argument and the resulting continuum equation are general; they hold for any non-negative, non-decreasing rate function $f$. Of course, if $f$ is complicated, so too will the continuum equation be (as in Appendix~\ref{app:nonlin}), and simulating the corresponding particle model will likely be preferable. As described in Section~\ref{sec:dimension}, we can use the continuum equation to fit the time constant $d$ with field data, which allows simulation timesteps to be translated into the timescale of the data. We emphasize that the scaling procedure both identifies a continuum model, as well as justifies the continuum model's approximation by simulations of the particle model, assuming $L$ is relatively large. The dimensionalization procedure of Section~\ref{sec:dimension} confirms that this condition will be satisfied in practice, as typical values for grain diameter and hillslope length give $L = 10^5$. 

We anticipate that the modeling approach described here will be particularly useful for long-timescale simulations and simulations of landscape relaxation in response to perturbations. As simulations of the particle model are easy to implement and computationally inexpensive, they could be used to evaluate the long-term impact of external drivers or could be incorporated as one component of a larger landscape model (e.g. hillslope with runoff into a river) while respecting modest computational resources. In addition, the simplicity of the particle model makes it possible to simulate the interaction of sophisticated perturbations, such as intermittent weather patterns or avalanching, with baseline geomorphic processes. Equipped with the dimensionalization procedure, these simulations can be informed by observations of individual grains and entire hillslopes, as well as stationary and perturbed hillslopes, and ultimately translated into concrete predictions.

\appendix

\section{Mathematical details}
\subsection{The product of one-parameter marginal distributions satisfies detailed balance.}\label{app:detailedbalance}
Following the argument of \citet{balazs2016product}, we show that the product distribution of Equation \ref{eq:dist} satisfies the detailed balance condition given in Equation \ref{eq:detailedbalance}, for bulk sites $i \neq 1,\,L$; the boundary cases follow from a similar argument.
\begin{linenomath*}\begin{align}
    p\,f\,(\omega (i))& \,{\mathds{P}_i}^{\theta_i} (\omega (i)) \,{\mathds{P}_{i+1}}^{\theta_{i+1}} (\omega (i+1)) \prod_{j \neq i,\,i+1} {\mathds{P}_j}^{\theta_j}\\[1em] 
    &= q \,f\,(\omega (i+1) + 1) \, {\mathds{P}_i}^{\theta_i} (\omega (i) - 1) \, {\mathds{P}_{i+1}}^{\theta_{i+1}} (\omega (i+1) + 1) \prod_{j \neq i,\,i+1} {\mathds{P}_j}^{\theta_j} \\[2em]
    p\,f\,(\omega (i))& \frac{e^{\theta_i \omega (i)}}{f\,(\omega (i))! \, Z(\theta_i)} \frac{e^{\theta_{i+1} \omega (i+1)}}{f\,(\omega (i+1))! \, Z(\theta_{i+1})}\\[1em]
    &= q \, f(\omega (i+1) + 1) \frac{e^{\theta_i (\omega (i) -1)}}{f\,(\omega (i) -1)! \, Z(\theta_i)} \frac{e^{\theta_{i+1} (\omega (i+1)+1)}}{f\,(\omega (i+1)+1)! \, Z(\theta_{i+1})}\\[2em]
    p\,f\,(\omega (i)) &= 
    q \, f(\omega (i+1) + 1) \frac{e^{\theta_{i+1}}}{f\,(\omega (i+1)+1)}\frac{f\,(\omega (i))}{e^{\theta_i}}\\[2em]
    p\,f\,(\omega (i)) &= q \, f\,(\omega (i)) e^{(\theta_{i+1} - \theta_i)}.
\end{align}\end{linenomath*} The last equation is satisfied when $\exp (\theta_{i+1} - \theta_i) = p/q$ and shows that the product distribution satisfies the bulk reversibility equations. %When applied to the boundary cases, this argument produces two additional constraints; namely, $\alpha(\omega (1)) = p f(\omega (1))$ and $\beta(\omega (L)) = q f(\omega (L))$.

\subsection{The expected occupancy for $f(\omega (i))  = \omega (i)$.}\label{app:density}

If $X_i$ are independent Poisson random variables with respective parameters $\lambda_i$ then, for $Y = \sum_{i=1}^n X_i$, the following argument shows $X_i | Y = k$ is binomially distributed with parameters $k$ and $\lambda_i / \sum_{j=1}^n \lambda_j$. $Y$ is the sum of independent Poisson random variables, so it is also Poisson and has parameter $\mu = \sum_{i=1}^m \lambda_i$. Call $Z_i = \sum_{j\neq i} X_j$, which is Poisson with parameter $\mu - \lambda_i$.
\begin{linenomath*}\begin{align}
    \mathds{P}(X_i = a | Y = k) &= \frac{\mathds{P}(X_i = a \cap Y = k)}{\mathds{P}(Y = k)}\\
    &= \frac{\mathds{P}(X_i = a)\cdot\mathds{P}(Z_i = k - a)}{\mathds{P}(Y=k)}\\
    &= \frac{{\lambda_i}^a e^{-\lambda_i}}{a!} \frac{{(\mu - \lambda_i)}^{k-a} e^{-{(\mu - \lambda_i)}}}{(k-a)!}\frac{k!}{\mu^k e^{-\mu}}\\
    &= \binom{k}{a}\Big(\frac{\lambda_i}{\mu}\Big)^a \Big(\frac{\mu - \lambda_i}{\mu}\Big)^{k-a}
\end{align}\end{linenomath*} where we used the independence of $X_i$ and $Z_i$ to get from the first line to the second.

Because the stationary distributions ${\mathds{P}_i}^{\theta_i}$ are Poisson when $f(\omega (i))  = \omega (i)$, we can apply this fact to Equation \ref{eq:density} as
\begin{linenomath*}\begin{equation}\label{eq:binomialdensity}
	{\rho (i)}^{\theta_i | H} = \mathds{E}^{\theta_i}\Big(\omega (i) \Big| \sum_{j=1}^L \omega (j) = H\Big) = \sum_{\omega (i)=0}^H \omega (i) \cdot {\mathds{P}_i}^{\theta_i} \Big(\omega (i) \Big| \sum_{j = 1}^L \omega (j) = H\Big). 
\end{equation}\end{linenomath*} We identify Equation \ref{eq:binomialdensity} as the mean of a binomial distribution with parameters $H$ and $e^{\theta_i}/\sum_{j=1}^L e^\theta_j$ to conclude
\begin{linenomath*}\begin{equation}
	{\rho (i)}^{\theta_i | H} = H\frac{e^{\theta_i}}{\sum_{j=1}^L e^{\theta_j}}.
\end{equation}\end{linenomath*}

\subsection{$\rho (i)$ is a strictly increasing function of $\theta_i$}\label{app:rho}

As $\rho (i)$ is an observable quantity, but $\theta_i$ is not, it is preferable that we parametrize expectations with $\rho (i)$ in the continuum limit. To do so, we need to show that their relation is invertible. It suffices for us to show that $\rho (i)$ is a strictly increasing function of $\theta_i$.
\begin{linenomath*}\begin{equation}
	{\rho (i)}^{\theta_i} = \mathds{E}^{\theta_i} \big( \omega (i) \big) = \sum_{k=0}^\infty \frac{k \cdot e^{{\theta_i} k}}{f\,(k)!\,Z({\theta_i})}
\end{equation}\end{linenomath*} and so
\begin{linenomath*}\begin{align}
	\frac{d}{d\theta}{\rho (i)}^{\theta_i} &= \sum_{k=0}^\infty \frac{k^2 \cdot e^{{\theta_i} k}}{f\,(k)!\,Z({\theta_i})} - \sum_{k=0}^\infty \frac{k \cdot e^{{\theta_i} k}}{f\,(k)!\,Z({\theta_i})} \cdot \cfrac{\frac{d}{d{\theta_i}}Z({\theta_i})}{Z({\theta_i})}\\
	&= \sum_{k=0}^\infty \frac{k^2 \cdot e^{{\theta_i} k}}{f\,(k)!\,Z({\theta_i})} - \Bigg( \sum_{k=0}^\infty \frac{k \cdot e^{{\theta_i} k}}{f\,(k)!\,Z({\theta_i})} \Bigg)^2 \\
	&= \mathds{E}^{\theta_i} \big( {\omega (i)}^2 \big) - \Big( \mathds{E}^{\theta_i} \big( \omega (i) \big) \Big)^2 > 0 \quad \forall \, \omega (i).
\end{align}\end{linenomath*} As ${\rho (i)}^{\theta_i}$ is a strictly increasing function of ${\theta_i}$, we can invert it to get ${\theta_i}({\rho (i)})$ and so can parametrize expectations in terms of an observable $\rho$.

\subsection{Solving the continuum equation}\label{sec:fokker}

We consider the setting of Section~\ref{sec:hydro} and, in particular, the continuum equation with Robin boundary conditions
\begin{linenomath*}\begin{align}\label{eq:hyd}
  \frac{\partial}{\partial t} \rho_t(x) &= \frac d2 \frac{\partial^2}{\partial x^2} G\bigl(\rho_t(x)\bigr)-2dE\frac{\partial}{\partial x} G\bigl(\rho_t(x)\bigr),\nonumber\\
  \frac{\partial}{\partial x} G\bigl(\rho_t(0)\bigr)&=4EG\bigl(\rho_t(0)\bigr),\\
  \frac{\partial}{\partial x} G\bigl(\rho_t(\ell)\bigr)&=4EG\bigl(\rho_t(\ell)\bigr).\nonumber
\end{align}\end{linenomath*}

\subsubsection{The linear case}\label{app:lin}
When the rates \(f\) are linear, \(G\) becomes the identity function and the above turns into the constant coefficient advection-diffusion equation
\begin{linenomath*}\begin{align}\label{eq:fp}
  \frac{\partial}{\partial t}\rho_t(x)&=\frac d2\frac{\partial^2}{\partial x^2}\rho_t(x)-2dE\frac{\partial}{\partial x}\rho_t(x),\nonumber\\
  \frac{\partial}{\partial x}\rho_t(0)&=4E\rho_t(0),\\
  \frac{\partial}{\partial x}\rho_t(\ell)&=4E\rho_t(\ell).\nonumber
\end{align}\end{linenomath*}
Notice that the time-stationary solution of \eqref{eq:fp} that we need is \(\rho(x)=\frac{4E h}{1-e^{4E\ell}}e^{4Ex}\). This is because the rescaled height profile then becomes
\begin{linenomath*}\begin{equation}
 h(x)=\lim_{L\to\infty}\frac1L\sum_{i=\lfloor xL\rfloor}^L\varrho_i=\lim_{L\to\infty}\sum_{i=\lfloor xL\rfloor}^L\rho\left(\frac iL\right)\frac1L=\int_x^\ell\rho(z)\,\text{d} z=\frac{h}{1-e^{4E\ell}}\bigl(e^{4Ex}-e^{4E\ell}\bigr)\label{eq:hx}
\end{equation}\end{linenomath*}
as needed for boundary conditions 0 at \(x=\ell\) and rescaled height \(h\) at \(x=0\). We now introduce the perturbation
\begin{linenomath*}\[
 \bar\rho_t(x)=\rho_t(x)-\rho(x)
\]\end{linenomath*}
and notice that this also satisfies \eqref{eq:fp}. However, it now makes physical sense to start with small initial data \(\bar\rho_0(x)\).

As \eqref{eq:fp} describes a drifting diffusion, it is natural to introduce
\begin{linenomath*}\[
 u_t(y)=\bar\rho_t(y+2dEt),\qquad-2dEt\le y\le1-2dEt.
\]\end{linenomath*}
Then
\begin{linenomath*}\begin{align*}
  \bar\rho_t(x)&=u_t(x-2dEt),&\qquad\frac{\partial}{\partial t}\bar\rho_t(x)&=\frac{\partial}{\partial t} u_t(x-2dEt)-2dE\frac{\partial}{\partial x} u_t(x-2dEt),\\
  \frac{\partial}{\partial x} \bar\rho_t(x)&=\frac{\partial}{\partial x} u_t(x-2dEt),&\qquad\frac{\partial^2}{\partial x^2} \bar\rho_t(x)&=\frac{\partial^2}{\partial x^2} u_t(x-2dEt),
\end{align*}\end{linenomath*}
and \eqref{eq:fp} becomes
\begin{linenomath*}\begin{align*}
  \frac{\partial}{\partial t} u_t(y)&=\frac d2\frac{\partial^2}{\partial y^2} u_t(y),\\
  \frac{\partial}{\partial y} u_t(-2dEt)&=4E u_t(-2dEt),\\
  \frac{\partial}{\partial y} u_t(1-2dEt)&=4E u_t(1-2dEt).
\end{align*}\end{linenomath*}
The first line is the ordinary heat equation, while the boundary conditions become rather unusual. As these are satisfied by \(u_t(y)\equiv0\), we expect that at least for times much smaller than \(\frac1{2dE}\) the boundary will not play a significant role in the solution if the initial condition \(u_0\) is small. Hence the solution should be close to
\begin{linenomath*}\begin{multline}
 u_t(y)=\frac1{\sqrt{2\pi dt}}\int_{-\infty}^\infty e^ {-\frac{(y-z)^2}{2dt}} u_0(z)\,\text{d} z,\qquad \text{or}\\ 
 \bar\rho_t(x)=u_t(x-2dEt)=\frac1{\sqrt{2\pi dt}}\int_{-\infty}^\infty e^{-\frac{(x-2dEt-z)^2}{2dt}}\bar\rho_0(z)\,\text{d} z.\label{eq:relax}
\end{multline}\end{linenomath*}

\subsubsection{The nonlinear case}\label{app:nonlin}
Here we consider a general but smooth \(G\) with derivative \(G'>0\) bounded away from zero in the relevant range of densities. \(G\) and \(G'\) are often not explicit but enjoy pleasant properties for particular models. The time-stationary solution of \eqref{eq:hyd} is \(G\bigl(\rho(x)\bigr)=c e^{4Ex}\) with a constant \(c\) that gives
\begin{linenomath*}\[
h=\int_0^1\rho(z)\,\text{d} z=\int_0^1G^{-1}\bigl(c e^{4Ez}\bigr)\,\text{d} z=\frac1{4E}\int_{G^{-1}(c)}^{G^{-1}(c e^{4E})}v\bigl(\ln G(v)\bigr)'\,\text{d} v.
\]\end{linenomath*}
Notice that this solves
\begin{linenomath*}\begin{align*}
  \frac12 \frac{\partial^2}{\partial x^2} G\bigl(\rho(x)\bigr)&=2E\frac{\partial}{\partial x} G\bigl(\rho(x)\bigr),\\
  \frac{\partial}{\partial x}G\bigl(\rho(0)\bigr)&=4EG\bigl(\rho(0)\bigr),\\
  \frac{\partial}{\partial x}G\bigl(\rho(\ell)\bigr)&=4EG\bigl(\rho(\ell)\bigr),
 \end{align*}\end{linenomath*}
that is
\begin{linenomath*}\begin{align}
  \frac12G''\bigl(\rho(x)\bigr)\Bigl(\frac{\partial}{\partial x}\rho(x)\Bigr)^2+\frac12G'\bigl(\rho(x)\bigr)\frac{\partial^2}{\partial x^2}\rho(x)&=2EG'\bigl(\rho(x)\bigr)\frac{\partial}{\partial x}\rho(x),\nonumber \\
  G'\bigl(\rho(0)\bigr)\frac{\partial}{\partial x}\rho(0)&=4EG\bigl(\rho(0)\bigr),\label{eq:stG}\\
  G'\bigl(\rho(\ell)\bigr)\frac{\partial}{\partial x}\rho(\ell)&=4EG\bigl(\rho(\ell)\bigr).\nonumber 
\end{align}\end{linenomath*}
As above, let
\begin{linenomath*}\[
 \bar\rho_t(x)=\rho_t(x)-\rho(x)=\rho_t(x)-G^{-1}\bigl(c e^{4Ex}\bigr).
\]\end{linenomath*}
Assuming this (and its derivatives) are small, we have
\begin{linenomath*}\begin{align*}
  \frac{\partial}{\partial t}\rho_t(x)&=\frac{\partial}{\partial t}\bar\rho_t(x),\\
  G\bigl(\rho_t(x)\bigr)&=G\bigl(\rho(x)\bigr)+G'\bigl(\rho(x)\bigr)\cdot\bar\rho_t(x)+\mathcal{O}\bigl(\bar\rho_t(x)\bigr)^2,\\
  \frac{\partial}{\partial x}G\bigl(\rho_t(x)\bigr)&=G'\bigl(\rho_t(x)\bigr)\frac{\partial}{\partial x}\rho_t(x)\\
  &=G'\bigl(\rho(x)\bigr)\Bigl(\frac{\partial}{\partial x}\bar\rho_t(x)+\frac{\partial}{\partial x}\rho(x)\Bigr)+G''\bigl(\rho(x)\bigr)\cdot\bar\rho_t(x)\cdot\Bigl(\frac{\partial}{\partial x}\bar\rho_t(x)+\frac{\partial}{\partial x}\rho(x)\Bigr)+\mathcal{O}\bigl(\bar\rho_t(x)\bigr)^2\\
  &=G'\bigl(\rho(x)\bigr)\Bigl(\frac{\partial}{\partial x}\bar\rho_t(x)+\frac{\partial}{\partial x}\rho(x)\Bigr)+G''\bigl(\rho(x)\bigr)\cdot\bar\rho_t(x)\cdot\frac{\partial}{\partial x}\rho(x)+\mathcal{O}\bigl(\bar\rho_t(x)\bigr)^2,\\
  \frac{\partial^2}{\partial x^2}G\bigl(\rho_t(x)\bigr)&=G''\bigl(\rho_t(x)\bigr)\Bigl(\frac{\partial}{\partial x}\rho_t(x)\Bigr)^2+G'\bigl(\rho_t(x)\bigr)\frac{\partial^2}{\partial x^2}\rho_t(x)\\
  &=G''\bigl(\rho(x)\bigr)\Bigl(\frac{\partial}{\partial x}\bar\rho_t(x)+\frac{\partial}{\partial x}\rho(x)\Bigr)^2+G'''\bigl(\rho(x)\bigr)\cdot\bar\rho_t(x)\cdot\Bigl(\frac{\partial}{\partial x}\bar\rho_t(x)+\frac{\partial}{\partial x}\rho(x)\Bigr)^2\\
  &\quad+G'\bigl(\rho(x)\bigr)\Bigl(\frac{\partial^2}{\partial x^2}\bar\rho_t(x)+\frac{\partial^2}{\partial x^2}\rho(x)\Bigr)+G''\bigl(\rho(x)\bigr)\cdot\bar\rho_t(x)\cdot\Bigl(\frac{\partial^2}{\partial x^2}\bar\rho_t(x)+\frac{\partial^2}{\partial x^2}\rho(x)\Bigr)+\mathcal{O}\bigl(\bar\rho_t(x)\bigr)^2\\
  &=G''\bigl(\rho(x)\bigr)\Bigl(\frac{\partial}{\partial x}\bar\rho_t(x)+\frac{\partial}{\partial x}\rho(x)\Bigr)^2+G'''\bigl(\rho(x)\bigr)\cdot\bar\rho_t(x)\cdot\Bigl(\frac{\partial}{\partial x}\rho(x)\Bigr)^2\\
  &\quad+G'\bigl(\rho(x)\bigr)\Bigl(\frac{\partial^2}{\partial x^2}\bar\rho_t(x)+\frac{\partial^2}{\partial x^2}\rho(x)\Bigr)+G''\bigl(\rho(x)\bigr)\cdot\bar\rho_t(x)\cdot\frac{\partial^2}{\partial x^2}\rho(x)+\mathcal{O}\bigl(\bar\rho_t(x)\bigr)^2.
 \end{align*}\end{linenomath*}
Combine this with \eqref{eq:hyd} and \eqref{eq:stG} to obtain
\begin{linenomath*}\begin{align*}
  \frac{\partial}{\partial t}\bar\rho_t(x)&=\frac d2G'\bigl(\rho(x)\bigr)\cdot\frac{\partial^2}{\partial x^2}\bar\rho_t(x)+\Bigl(dG''\bigl(\rho(x)\bigr)\frac{\partial}{\partial x}\rho(x)-2dEG'\bigl(\rho(x)\bigr)\Bigr)\cdot\frac{\partial}{\partial x}\bar\rho_t(x)\\
  &\quad+\Bigl(\frac d2G'''\bigl(\rho(x)\bigr)\Bigl(\frac{\partial}{\partial x}\rho(x)\Bigr)^2+\frac d2G''\bigl(\rho(x)\bigr)\frac{\partial^2}{\partial x^2}\rho(x)-2dEG''\bigl(\rho(x)\bigr)\frac{\partial}{\partial x}\rho(x)\Bigr)\cdot\bar\rho_t(x)\\
  &\quad+\mathcal{O}\bigl(\bar\rho_t(x)\bigr)^2,\\
  G'\bigl(\rho(0)\bigr)\cdot\frac{\partial}{\partial x}\bar\rho_t(0)&=\Bigl(4EG'\bigl(\rho(0)\bigr)-G''\bigl(\rho(0)\bigr)\frac{\partial}{\partial x}\rho(0)\Bigr)\cdot\bar\rho_t(0)+\mathcal{O}\bigl(\bar\rho_t(0)\bigr)^2,\\
  G'\bigl(\rho(\ell)\bigr)\cdot\frac{\partial}{\partial x}\bar\rho_t(\ell)&=\Bigl(4EG'\bigl(\rho(\ell)\bigr)-G''\bigl(\rho(\ell)\bigr)\frac{\partial}{\partial x}\rho(\ell)\Bigr)\cdot\bar\rho_t(\ell)+\mathcal{O}\bigl(\bar\rho_t(\ell)\bigr)^2.\\
 \end{align*}\end{linenomath*}
Neglecting error terms, the result is a linear equation, which may be solved numerically and used to fit the time constant \(d\).

\subsection{Estimating typical distances traveled by particles}\label{sec:dists}

We begin with a disclaimer: this section is \textit{not} part of the core argument connecting the particle model of Section~\ref{sec:model} to the continuum hillslope description of Section~\ref{sec:hydro}. The contents of this section are instead intended as an example of how one might infer average distances traveled by hillslope particles; we cannot calculate this directly, as the ``particles'' of our model are units of gradient, not hillslope particles. To overcome this barrier, we settle for an intuitive, mean-field argument. Note that no rescaling is involved, since one step of a grain is not imagined on scales comparable to the size of the hillslope. We therefore consider the slope \(\varrho=\mathbb{E} \omega_i\) a constant parameter that changes as we look at different parts of the hill.

Consider, for the sake of argument, a medium flowing over the hillslope, which lifts, carries, and deposits hillslope particles, building up the heights \(h_i\). It is assumed that this medium flows at velocity \(v(\varrho)\) [units of \(i\) / model time \(\tau\) units] and that it tracks with particle deposition and removal, which happens at an average rate of $p e^{\theta(\varrho)}$. In other words, it takes an average time of $1/(p e^{\theta (\varrho)})$ for the flow to move one unit of distance (one site to the next), and so we write \begin{linenomath*}\[ v (\varrho) = p e^{\theta (\varrho)}.\]\end{linenomath*} Notice that this is an increasing function of the slope $\varrho$ as one would expect, and that under our scaling $p$ is close to $1/2$, which we will substitute.

We assume that a given grain spends an average time $\tau_0 (\varrho)$ transported by the flow before depositing . The function $\tau_0$ is an input of the model and might be chosen as a constant or, perhaps more naturally, as an increasing function of $\varrho$. This gives a deposition rate of $1/\tau_0 (\varrho)$ and so the average distance traveled is \begin{linenomath*}\[ D (\varrho) = v (\varrho) \cdot \tau_0 (\varrho) = \frac{1}{2} e^{\theta (\varrho)} \cdot \tau_0 (\varrho). \]\end{linenomath*}

We assume that an average number $n (\varrho)$ of grains are carried by the flow per (microscopic) site (of the particle model). As over sufficiently long timescales the hillslope does not grow or vanish, the average flux \(\Psi\) of carried grains, \(v(\varrho)\cdot n(\varrho)\) is conserved across the hillslope, from which we can assert
\begin{linenomath*}\[
 n(\varrho)=\frac\Psi{v(\varrho)} = 2 \Psi e^{-\theta (\varrho)},
\]\end{linenomath*} a decreasing function of the slope $\varrho$. As each particle settles at rate $1/\tau_0 (\varrho)$, the total rate at which particles are deposited at an individual site is \begin{linenomath*}\[ \frac{n(\varrho)}{\tau_0 (\varrho)} = \frac{2\Psi e^{-\theta (\varrho)}}{\tau_0 (\varrho)}. \]\end{linenomath*}

An essential feature of this model is to distinguish between a particle depositing on the hillslope and growth of a column in the gradient particle model. As the latter happens at an average rate of $\frac{1}{2} e^{\theta (\varrho)}$, every column-raising event of the gradient process is considered a deposition event for the hillslope as well with probability \begin{linenomath*}\[ \frac{4\Psi e^{-2\theta (\varrho )}}{\tau_0 (\varrho)},\]\end{linenomath*} which must therefore be less than 1. Due to reversibility, we have the same rates and probabilities for entrainment.

A given particle takes part in a column growth event at average rate \begin{linenomath*}\[ \frac{e^{\theta(\varrho)}}{2n(\varrho)} = \frac{e^{2\theta (\varrho)}}{4\Psi},\]\end{linenomath*} an increasing function of slope. Multiplying this with the probability from the previous line recovers $1/\tau_0 (\varrho)$ as the deposition rate.

To conclude, we have the folllowing examples of average distance traveled:
\begin{linenomath*}\[
 D(\varrho)=\left\{
  \begin{aligned}
   & \frac{1}{2} \varrho \, \tau_0 (\varrho) &&\text{for linear rate,}\\
   &\frac{1}{2} \frac{\varrho}{1+\varrho} \, \tau_0 (\varrho) &&\text{for constant rate.}
  \end{aligned}
 \right.
\]\end{linenomath*}

\acknowledgments
The authors acknowledge the generous support of the Marshall Scholarship (JC) and Hungarian Scientific Research Fund (OTKA/NKFIH) grant K109684 (MB). 

\listofchanges
%%%

%\bibliography{references}
%\addcontentsline{toc}{chapter}{References}
%\bibliographystyle{plainnat}

\end{document}